\documentclass[prb,twocolumn,showpacs,preprintnumbers,amsmath,amssymb]{revtex4}

\usepackage{graphicx}
\usepackage{dcolumn}
\usepackage{bm}
\usepackage{eucal}

\begin{document}
\title{Chiral spin-wave excitations of the spin-5/2 trimers\\ in
the langasite compound Ba$_3$NbFe$_3$Si$_2$O$_{14}$}
\author{Jens Jensen}
\affiliation{Niels Bohr Institute, Universitetsparken 5, 2100
Copenhagen, Denmark}
\date{\today}
\begin{abstract} The inelastic scattering of neutrons from
magnetic excitations in the antiferromagnetic phase of the langasite
compound Ba$_3$NbFe$_3$Si$_2$O$_{14}$ is analyzed theoretically. In
the calculations presented, the strongly coupled spin-5/2 Fe
triangles are accounted for as trimerized units. The weaker
interactions between the trimers are included within the
mean-field/random-phase approximation. The theory is compared with
linear spin-wave theory, and a model is developed which leads to good
agreement with the published results from unpolarized and polarized
neutron-scattering experiments.
\end{abstract}
 \maketitle
\section{Introduction}
\begin{figure}[t]
\includegraphics[width=0.9\linewidth]{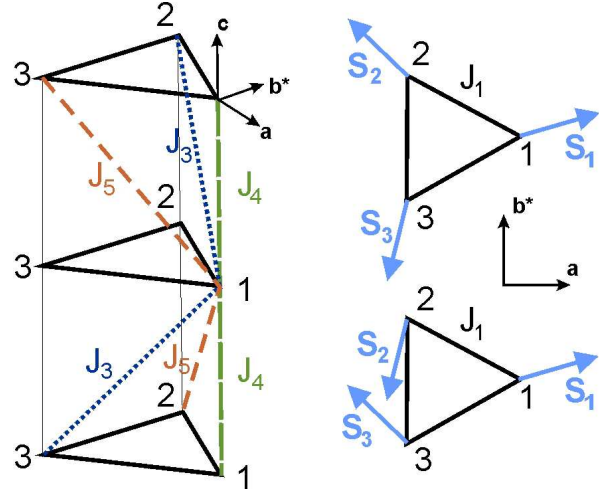}
\caption{(Color online) Fe triangles are stacked on top of each other along the $c$
axis in the Ba$_3$NbFe$_3$Si$_2$O$_{14}$ crystal. The line parallel to the
$c$ axis through the center of the triangles is a
three-fold symmetry axis, and the three altitudes of one triangle are
two-fold symmetry axes. The oxygen ions are placed in different positions along
the paths determining the interplanar exchange constants
$J_3^{}$, $J_4^{}$, and $J_5^{}$. The shortest super-superexchange path
(and bond angles closest to $\pi$) defines\cite{Marty} the structural chirality
$\epsilon_T^{}=\pm1$.
If this path is the one determining $J_5^{}$ as defined in the figure, the
corresponding chirality would be $\epsilon_T^{}=-1$, whereas $\epsilon_T^{}=+1$
if this path is the one leading to the exchange interaction $J_3^{}$.
The right part of the figure shows the
ordered spins in one of the triangles, where the moments rotate the angle
$\gamma=\epsilon_\gamma^{}(2\pi/3)$ for increasing site index. The upper triangle
shows the case where $\epsilon_\gamma^{}=+1$, whereas the
orientation of the ordered spins in the lower triangle corresponds to
$\epsilon_\gamma^{}=-1$.} \label{f1}
\end{figure}

\begin{figure}[t]
\includegraphics[width=0.6\linewidth]{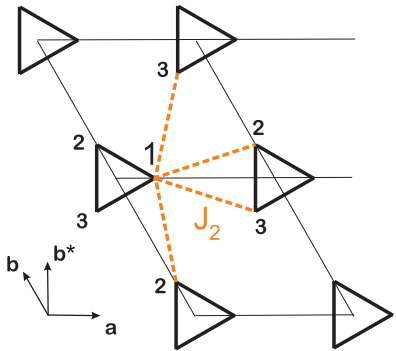}
\caption{(Color online) The triangles within an $ab$ plane with a definition of the
exchange parameter $J_2^{}$.}
\label{f2}
\end{figure}

The langasite compound Ba$_3$NbFe$_3$Si$_2$O$_{14}$ belongs to the
space group $P321$ (space group number 150), where triangles of
Fe$^{3+}$ ($L=0$, $S=5/2$) ions are placed in a lattice as indicated
in Figs.\ \ref{f1} and \ref{f2}. The space group contains no improper
symmetry elements and the other ions may be arranged in two different
mirrored ways as specified by the sign of the structural chirality
factor $\epsilon_T^{}=\pm1$. The Hamiltonian for the $S=5/2$ spins of
the Fe ions is assumed to be
\begin{equation}\label{e0}
{\cal H}=\frac{1}{2}\sum_{i,\xi}\sum_{j,\eta
}J_{\xi\eta}^{}(ij)\,\mathbf{S}_\xi(i)
\cdot\mathbf{S}_\eta(j)+\sum_i{\cal H}_T^{}(i),
\end{equation}
where $i$ and $j$ are the triangle numbers, and $\xi$ and
$\eta=1,2,3$ denote the different spins in each triangle, as defined
in Fig.\ \ref{f1}. The Hamiltonian for the isolated triangles is
determined in terms of the intra-triangle interaction $J_1^{}>0$ and
an anisotropy term assumed to be a Dzyaloshinsky--Moriya (DM)
interaction:
\begin{eqnarray}\label{e01}
{\cal H}_T^{}&=&
J_1^{}\left(\mathbf{S}_1^{}\cdot\mathbf{S}_2^{}
+\mathbf{S}_2^{}\cdot\mathbf{S}_3^{}+\mathbf{S}_3^{}\cdot\mathbf{S}_1^{}
\right)\\ &&+D_c^{}\left(\mathbf{S}_1^{}\times\mathbf{S}_2^{}
+\mathbf{S}_2^{}\times\mathbf{S}_3^{}+\mathbf{S}_3^{}\times\mathbf{S}_1^{}\right)
\cdot\mathbf{\hat{c}}\nonumber\,.
\end{eqnarray}

The system is antiferromagnetically ordered below $T_N^{}=27$
K.\cite{Marty,MartyJMMM} The moments are confined to lie in the $ab$
plane, which property is in accordance with the anisotropy introduced
in Eq.\ (\ref{e01}) (independent of the sign of $D_c^{}$). All
triangles in a certain $ab$ plane are identical, and the three spins,
$\langle\mathbf{S}_1^{i}\rangle$, $\langle\mathbf{S}_2^{i}\rangle$,
and $\langle\mathbf{S}_3^{i}\rangle$ in the $i$th triangle are making
an angle of $\gamma=\epsilon_\gamma^{}(2\pi/3)$ with each other so
that $\langle\mathbf{S}_1^{i}\rangle+\langle\mathbf{S}_2^{i}\rangle+
\langle\mathbf{S}_3^{i}\rangle=\mathbf{0}$, where
$\epsilon_\gamma^{}=\pm1$ defined in Fig.\ \ref{f1} denotes the two
possible orientations of the ordered spin triangles. The moments
along a line parallel to the $c$ axis rotate the angle
$\phi=\mathbf{Q}\cdot\mathbf{c}\simeq\epsilon_H^{}(2\pi/7)$ from one
layer to the next along the $c$ axis, where $\epsilon_H^{}=\pm1$
denotes the chirality/helicity of the helically ordered moments.

The magnetic excitations at low temperatures in the helically ordered
phase have been studied experimentally by two independent groups:
Stock {\it et al.}\cite{Stock} determined the dispersion relations of
the spin waves propagating in the $a^\ast c^\ast$ plane from
inelastic scattering of unpolarized neutrons, whereas Loire {\it et
al.}\cite{Loire} investigated the excitations by doing inelastic
scattering experiments in the $b^\ast c^\ast$ plane with both
unpolarized and polarized neutrons. Loire {\it et al.}\cite{Loire}
carried through a linear spin-wave analysis of their results from
which they concluded that the moments were all ordered in righthanded
helices, corresponding to $\epsilon_H^{}=+1$, in the
$\epsilon_T^{}=-1$ enantiopure crystal they were investigating.

The spin system in Ba$_3$NbFe$_3$Si$_2$O$_{14}$ is relatively
strongly frustrated, and the validity of the mean-field (MF) and the
random-phase approximation (RPA), on which the linear spin-wave
theory is based, may be questioned. The dominant cause for
frustration is the strong interaction $J_1^{}$ between the three
spins in the Fe triangles. In the present paper I have improved the
theoretical description of the system by treating the spin triangles
as trimerized units placed in a mean-field due to the interactions
between neighboring trimers. The properties of the trimers are
determined accurately by a numerical diagonalization of the
mean-field trimer Hamiltonian, and the corresponding spectrum of
collective trimer excitations is calculated by the use of standard
RPA numerical techniques.\cite{REM,JJ} Before the numerical analysis
is presented in Sect.\ III, it is instructive to consider the
approximate but analytical behavior of the spin system as done in
Sect.\ II. The spin-wave theory applied in this section is not the
standard one based on the Holstein-Primakoff transformation, but it
is straightforward to show that the present RPA approach and the
standard spin-wave theory are equivalent to leading order (see for
instance Ref.\ \onlinecite{REM}). In Sect.\ IV, the cross section for
polarized neutrons is calculated numerically, and the results are
compared with the experimental results obtained by Loire {\it et
al.}\cite{Loire} The conclusions are presented in Sect.\ V.

\section{Linear spin-wave theory}
A coordinate system is assigned to each spin, defined so that the
local $z$ axis is along the direction of the ordered moment, and the
$y$ axis is along the $c$ axis and common for all coordinate systems.
In the spin-wave limit, where the spins are fully polarized, the
ground state of the $i$th triangle is $|000\rangle$. The states are
defined by $\hat{S}_{\xi z}^{}|n_{1}^{}n_{2}^{}n_{3}^{}\rangle=
(n_{\xi}^{}-S)|n_{1}^{}n_{2}^{}n_{3}^{}\rangle$, where $\xi=1,2,3$.
The interactions between the different triangles lead to a mean field
acting only on the $z$ components (neglecting any anisotropy within
the $ab$ plane):
\begin{equation}\label{e1}
{\cal H}_{\text{MF}}^{}=-h\left(S_{1z}^{}+S_{2z}^{}+S_{3z}^{}\right)
+{\cal H}_T^{}\,.
\end{equation}
In this section it is assumed that $h\gg J_1^{}$ and only terms to
first order in $J_1^{}/h$ and $D_c^{}/h$ are included. The
intra-triangle interactions imply that the ground state $|g\rangle$
is modified into
\begin{eqnarray}\label{e2}
|g\rangle&=&|000\rangle
+\lambda\left(|110\rangle+|101\rangle+|011\rangle\right),\nonumber\\
\lambda&=&\frac{3SJ_1^{}-2SD_c^{}\sin\gamma}{8h}.
\end{eqnarray}
The first-order modification of the ground state does not affect the
ground-state energy to leading order, and it is
\begin{equation}\label{e3}
E(g)= -3Sh-{\textstyle\frac{3}{2}}S^2J_1^{}+3S^2D_c^{}\sin\gamma\,.
\end{equation}
The presence of the DM anisotropy implies a specific sign for
$\gamma$. The product $D_c^{}\sin\gamma$ has to be negative, and the
orientation of the ordered spin triangles is determined by the sign
of $D_c^{}$ according to $\epsilon_\gamma^{}=-\mbox{sign}(D_c^{})$.

In terms of the interaction parameters defined in Figs.\ \ref{f1} and
\ref{f2}, the exchange field is:
\begin{eqnarray}\label{e4}
h=-2S\left[\,2J_2^{}\cos\gamma\right.+J_4^{}\!\!\!&&\!\!\!\cos\phi\nonumber\\
+J_3^{}\cos(\phi+\gamma)\!\!\!&&\!\!\!+
\left.J_5^{}\cos(\phi-\gamma)\right].
\end{eqnarray}
$\phi$ is the angle of rotation of $\mathbf{S}_\alpha^{}$ within the
plane, when going from one $ab$ plane to the next in the positive $c$
direction. The value of $\phi$ is found by maximizing $h$ (minimizing
the ground state energy) and is determined by
\begin{equation}\label{e5}
\tan\phi=R\sin\gamma,\quad R=\frac{2(J_5-J_3)}{2J_4-J_3-J_5}\,.
\end{equation}
It is remarkable that a difference between $J_3^{}$ and $J_5^{}$
implies that $\phi$ becomes non-zero with no need for an interaction
between next-nearest neighboring spins along the $c$ axis.
Experimentally,\cite{MartyJMMM} it is found that $|\phi|$ is close to
$2\pi/7$. The relation also shows that the magnetic helicity and the
orientation of the ordered spin triangles are intimately related,
that
\begin{equation}\label{en}
\epsilon_H^{}=\mbox{sign}(R)\,\epsilon_\gamma^{}\,.
\end{equation}

The three lowest excited states are (approximately)
\begin{eqnarray}\label{e6}
|a\rangle&=&\left(|100\rangle+w|010\rangle+w^2|001\rangle\right)/\sqrt{3}\nonumber\\
|b\rangle&=&\left(|100\rangle+w^2|010\rangle+w|001\rangle\right)/\sqrt{3}\\
|c\rangle&=&\left(|100\rangle+|010\rangle+|001\rangle\right)/\sqrt{3}\,,\nonumber
\end{eqnarray}
where $w=e^{i2\pi/3}$. The corresponding excitation energies,
$\Delta_\alpha^{}=E(\alpha)-E(g)$, are
\begin{eqnarray}\label{e7}
&&\Delta_a^{}=\Delta_b^{}=h+{\textstyle\frac{3}{4}}SJ_1^{}-
{\textstyle\frac{5}{2}}SD_c^{}\sin\gamma\nonumber\\
&&\Delta_c^{}=h+{\textstyle\frac{3}{2}}SJ_1^{}-SD_c^{}\sin\gamma\,.
\end{eqnarray}
The collective excitations of this system may be derived by
introducing the ``boson-like'' (creation) operators for the $i$th
triangle: $\alpha_i^+|g\rangle=|\alpha\rangle$, where $\alpha=a$,
$b$, or $c$. Neglecting the interactions between the triangles and
introducing the Fourier transforms of the operators, the
single-trimer MF Hamiltonian may be written:
\begin{equation}\label{e8}
\mbox{}\!\!{\cal H}_{\text{MF}}^{}=E(g)+\frac{1}{N}\sum\left[\Delta_a^{}a_{\mathbf{q}}^+a_{\mathbf{q}}^{}
+\Delta_b^{}b_{\mathbf{q}}^+b_{\mathbf{q}}^{}+
\Delta_c^{}c_{\mathbf{q}}^+c_{\mathbf{q}}^{}\right].
\end{equation}
Within the subspace of the four lowest spin states for the $i$th triangle,
we may write
\begin{eqnarray}\label{e9}
S_{1x}^{}\!\!\!&=&\!\!\!\left[m_x^{c}(c_i^{}+c_i^+)+m_x^{a}(a_i^{}+a_i^+
+b_i^{}+b_i^+)\right]\sqrt{\frac{S}{6}}\qquad\\
S_{1y}^{}\!\!\!&=&\!\!\!i\left[m_y^{c}(c_i^{}-c_i^+)+m_y^{a}(a_i^{}-a_i^+
+b_i^{}-b_i^+)\right]\sqrt{\frac{S}{6}}\,,\qquad\nonumber
\end{eqnarray}
and similarly for the other spin components, except that the
$(a_i^{},b_i^+)$ and $(a_i^{+},b_i^{})$ terms are multiplied,
respectively, by $w$ and $w^2$ in the expressions for the spin
components of $\mathbf{S}_2^{}$, and by $w^2$ and $w$ in the spin
components of $\mathbf{S}_3^{}$. To first order in $J_1^{}/h$, the
relative matrix elements are
\begin{eqnarray}\label{e10}
&m_x^c=1+2\lambda,\qquad &m_y^c=1-2\lambda\nonumber\\
&m_x^a=1-\lambda,\phantom{2}\qquad &m_y^a=1+\lambda\,,
\end{eqnarray}
in terms of the mixing parameter $\lambda$ defined in (\ref{e2}).
Next step is to substitute these expressions for the spin components
in the exchange Hamiltonian, which leads to an effective Hamiltonian
quadratic in the excitations operators. Introducing the Fourier
transforms of the operators, the Hamiltonian stays diagonal with
respect to the $(c_{\mathbf{q}}^{},c_{\mathbf{q}}^{+})$ operators,
when $\mathbf{q}$ is parallel to the $c$ axis . We shall concentrate
on this case in the following, i.e.\ that
$\mathbf{q}\cdot\mathbf{c}=qc$. In the zero temperature limit, the
only non-zero commutator relations are
$[\alpha_{\mathbf{q}}^{},\alpha_{\mathbf{q'}}^{+}]\approx
\delta_{\mathbf{q}\mathbf{q'}}^{}$, where $\alpha=a$, $b$, or $c$.
From the equations of motion, the energy squared of the $c$-mode
excitations propagating along the $c$ axis is then found to be
\begin{eqnarray}\label{e11}
&&E_c^2(q)=\left[\Delta_c^{}+S(m_x^c)^2\left\{\left[J(Q)+2J_2^{}\right]
\cos(qc)-2J_2^{}\right\}\right]\nonumber\\
&&\times\left[\Delta_c^{}+S(m_y^c)^2\left\{\left[J(0)-4J_2^{}\right]
\cos(qc)+4J_2^{}\right\}\right],
\end{eqnarray}
where the two exchange parameters are
\begin{eqnarray}\label{e12}
J(Q)\!\!&=&\!\!2\left[J_4^{}\cos\phi+J_3^{}\cos(\phi+\gamma)+
J_5^{}\cos(\phi-\gamma)-J_2^{}\right]\nonumber\\
J(0)\!&=&\!\!2\left[2J_2^{}+J_3^{}+J_4^{}+J_5^{}\right].
\end{eqnarray}
$J(Q)$ is also the parameter determining the exchange field,
$h=-SJ(Q)$. The remaining part of the excitation Hamiltonian leads to
two $w$ modes with mixed $|a\rangle$ and $|b\rangle$ state
characters, and the squared energies of these modes are
\begin{eqnarray}\label{e13}
E_{w1}^2(q)\!\!&=&\!\!\left[\Delta_a^{}+S(m_x^a)^2{\cal J}_x^{}(q)\right]
\left[\Delta_a^{}+S(m_y^a)^2{\cal J}_y^{}(q)\right]\nonumber\\
E_{w2}^2(q)\!\!&=&\!\!\left[\Delta_a^{}+S(m_x^a)^2{\cal J}_x^{}(-q)\right]
\left[\Delta_a^{}+S(m_y^a)^2{\cal J}_y^{}(-q)\right],\nonumber\\
\end{eqnarray}
where
\begin{eqnarray}\label{e14}
{\cal J}_x^{}(q)&=&2J_4^{}\cos\phi\cos(qc)+2J_3^{}\cos(\phi+\gamma)
\cos(qc+\gamma)\nonumber\\ &&{}\qquad+2J_5^{}\cos(\phi-\gamma)\cos(qc-\gamma)
+J_2^{}\nonumber\\
{\cal J}_y^{}(q)&=&2J_4^{}\cos(qc)+2J_3^{}
\cos(qc+\gamma)\\ &&{}\qquad+2J_5^{}\cos(qc-\gamma)-2J_2^{}\,.\nonumber
\end{eqnarray}
In general ${\cal J}_{x,y}^{}(-q)\ne{\cal J}_{x,y}^{}(q)$, which
means that the two modes are not symmetric around $q=0$, instead we
have that $E_{w1}^{}(q)=E_{w2}^{}(-q)$.

Introducing the expressions for $\Delta_\alpha^{}$ and the relative
matrix elements into Eqs.\ (\ref{e11}) and (\ref{e13}), the squared
excitations energies are, to leading order in $\lambda$,
\begin{eqnarray}\label{e15}
&&E_c^2(q)=S^2\left[1-\cos(qc)\right]\left[-J(Q)-2J_2^{}\right]\\
&&\!\!\!\!\!\!\times\left[3J_1^{}-2D_c^{}\sin\gamma-J(Q)+4J_2^{}+
\left\{J(0)-4J_2^{}\right\}\cos(qc)\right]\nonumber
\end{eqnarray}
and
\begin{eqnarray}\label{e16}
&&E_w^2(q)=S^2\left[{\cal J}_y^{}(q)-J(Q)-2D_c^{}\sin\gamma\right]\\
&&{}\qquad\qquad\times\left[{\textstyle\frac{3}{2}}J_1^{}-3D_c^{}\sin\gamma+
{\cal J}_x^{}(q)-J(Q)\right].\nonumber
\end{eqnarray}
Equation (\ref{e15}) shows that the energy of the $c$ mode
$E_c^{}(q)$ vanishes linearly with $q$, when $q\to0$. This Goldstone
mode, which appears because the rotational symmetry in spin space
around the $c$ axis is broken in the ordered phase, reflects that it
costs no energy to rotate the ordered structure around the $c$ axis.
In the case of a $c$ excitation propagating along the $c$ axis, all
the $c$ components of the spins in a certain $ab$ plane are moving in
phase, whereas the $w$ modes imply that the plane of the spins in a
certain triangle is oscillating out of the $ab$ plane, and this
oscillation is affected by the DM anisotropy. Because of this
anisotropy, the $w$ modes show an energy gap at $q=\pm Q$
\begin{equation}\label{e17}
E_{w}^{}(Q)\simeq S\left[-D_c^{}\sin\gamma\{3J_1^{}+
2{\cal J}_x^{}(\phi/c)-2J(Q)\}\right]_{}^{1/2}.
\end{equation}

The transverse $(x,y)$ components of the spins are defined in a
coordinate system for which the $x$ axis rotates the angle $\phi$
around the $y$ or $c$ axis from one $ab$ plane to the next. In a
neutron experiment with the scattering vector $\mathbf{k}$ along the
$c$ axis, modulus a reciprocal lattice vector, the excitations
detected are those described above at $q=k$, when the scattering
derives from the $y$ or $c$ component of the spins. If the scattering
is instead due to the $x$ component, the experiment detects the
excitations at $q=k+Q$ and $q=k-Q$. The $c$ mode behaves similarly to
the spin waves in a simple helix, where the $c$ component reflects
the branch for which the Goldstone mode starts out from $k=0$,
whereas an $ab$ scattering vector component reflects the two
branches, where the Goldstone modes emerge from the two magnetic
Bragg peaks at $k=\pm Q$. The situation is different for the $w$
modes. The excitations detected by the $c$ component are those
determined above at $q=k$, i.e.\ a $w1/w2$ mode emerging (subjected
to a small gap) from the Bragg point at $+Q$ and a $w2/w1$ mode
emerging from the other Bragg point at $-Q$. When the $w$ modes are
detected via the component in the $ab$ plane, the wave numbers of the
$c$-component branches are translated by $Q$ or $-Q$. Two of those
become pseudo Goldstone modes starting out from $k=0$, whereas the
two other branches are placed with their starting points at $k=2Q$
and $k=-2Q$. The intensities of the extra $\pm2Q$ branches are,
however, always going to be weak.

Although the $c$ mode shows similarities with the spin waves in a
simple helix there is one important difference, namely that the cross
section of the $c$ mode vanishes when the total scattering vector is
parallel to the $c$ axis. In this case the excitations are detected
exclusively via the spin components perpendicular to the $c$ axis.
For the $c$ modes propagating along the $c$ axis all the locally
defined spin components within one $ab$ plane oscillate in phase, and
this means that the sum of the $ab$ components for a spin-triangle
stays zero during the $c$-mode oscillations. The two $w$ branches
starting out from $\pm2Q$ have no cross section either, which leaves
the two pseudo Goldstone $w$ branches starting out from $k=0$ to be
the only ones appearing in a scan along $(0,0,q)$. Such a scan would,
effectively, only show a single spin-wave branch (see Fig.\ \ref{f7}
in the next section), in strong contrast to that expected in the case
of a simple helically ordered system. In general, ${\cal
J}_y^{}(q)={\cal J}_y^{}(2\phi/c-q)$ whereas ${\cal
J}_x^{}(q)\ne{\cal J}_x^{}(2\phi/c-q)$, which means that there are
actually two branches in the plot along $(0,0,q)$ shown in Fig.\
\ref{f7}, but the energy difference is small -- about 0.06 meV at
maximum around $q=1.25$.

In addition to the DM anisotropy included in (\ref{e1}), the symmetry
of the system also allows the following term\cite{Moriya}
\begin{equation}\label{e18}
D_{ab}^{}\left(
\mathbf{\hat{r}}_{12}^{}\cdot\mathbf{S}_1\times\mathbf{S}_2^{}+
\mathbf{\hat{r}}_{23}^{}\cdot\mathbf{S}_2\times\mathbf{S}_3^{}+
\mathbf{\hat{r}}_{31}^{}\cdot\mathbf{S}_3\times\mathbf{S}_1^{}\right),
\end{equation}
but this interaction does not affect any of the quantities considered
above to leading order in $D_{ab}^{}/h$. The numerical analysis
presented below shows that this coupling may weakly perturb the
ordered structure by inducing a small oscillating $c$ component on
top of the spiraling ordered moments lying in the $ab$ plane. The
ordered $c$ components have a constant magnitude within an $ab$
plane, which magnitude varies sinusoidally along the $c$ axis with
the same period as the helix. Using a $D_{ab}^{}$ with the same
magnitude as $D_c^{}=0.0038$ meV considered in the next section, the
scattering intensity due to the oscillating $c$-axis moment is found
to be a factor of $10^6$ smaller than the intensity due to the $a$ or
$b$ component. Hence, we may safely neglect any influences from
$D_{ab}^{}$. Finally, I may add that the simple anisotropy term
$D_z^{}(S_{1z}^2+S_{1z}^2+S_{1z}^2)$, with $D_z^{}\simeq0.005$ meV,
may replace $D_c^{}$ in the explanation for the confinement of the
moments to the $ab$ plane and for the presence of the energy gap
$E_w^{}(Q)$. In contrast to the DM term, the $D_z^{}$ anisotropy
would affect the susceptibility components in the paramagnetic phase,
but the value of $D_z^{}$ indicated by the energy gap $E_w^{}(Q)$ is
too small to make any observable difference. However, this
interaction has no influence on the choice of sign for $\gamma$, and
is therefore unable to explain why the system prefer one of the two
choices as found experimentally.\cite{Marty}

\begin{figure}[t]
\includegraphics[width=0.85\linewidth]{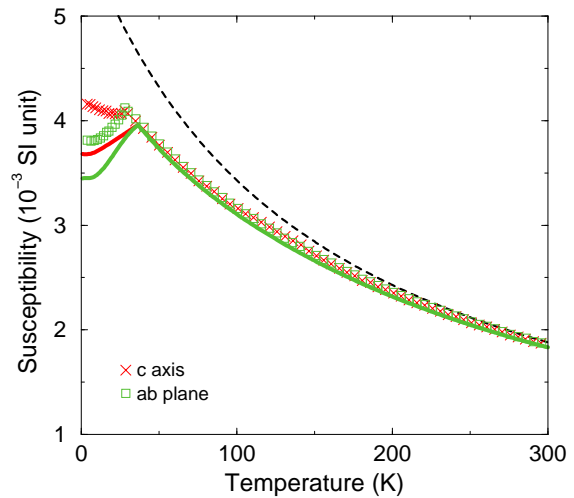}
\caption{(Color online) The calculated susceptibility components as functions of temperature
compared with the experimental results of Marty {\it et al.}\cite{Karol} The red
line and the red crosses denote, respectively, calculated and experimental results,
when the field is applied along the $c$ axis. The green symbols denote the results
obtained when the field is applied perpendicular to the $c$ axis. The dashed line
shows the MF behavior for the corresponding simple $S=5/2$ system. The difference
between the dashed and solid lines indicates the importance of correlation between
the three spins in the triangle clusters.}
\label{f3}
\end{figure}

\section{Numerical model calculations}
In the numerical analysis, the cluster MF Hamiltonian (\ref{e1}) is
diagonalized precisely and the excitation spectra are calculated
without making any further approximations than the basic random-phase
approximation, see Refs.\ \onlinecite{REM,JJ}. The best MF/RPA trimer
model obtained from fitting the susceptibility and excitation data
obtained by Stock {\it et al.}\cite{Stock} is (in units of meV):
\begin{eqnarray}\label{e19}
&&\!\!\!\!\!\!J_1^{}=1.25,~J_2^{}=0.2,~J_3^{}=0.1,~J_4^{}=0.064,~J_5^{}=0.29
\nonumber\\&&\!\!\!\!\!\!D_c^{}=0.0038.
\end{eqnarray}
The anisotropy is very small, and I have chosen the sign to be
positive. This means that $\gamma=-2\pi/3$, and introducing the
exchange parameter given above into the equilibrium condition
(\ref{e5}) leads to $\tan\phi=1.257$ or $\phi=2\pi/7$.

\begin{figure}[t]
\includegraphics[width=0.9\linewidth]{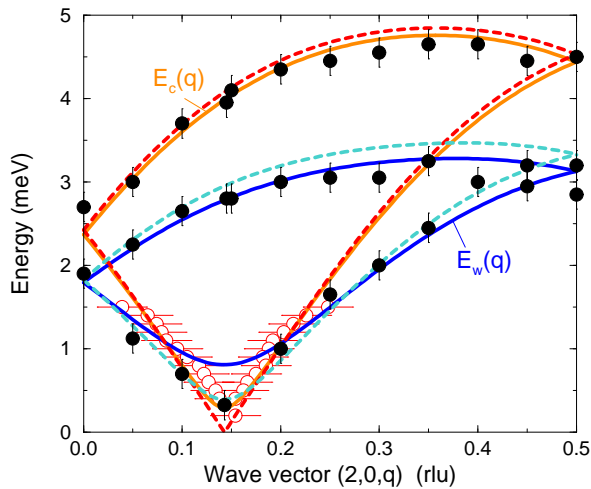}
\caption{(Color online) The calculated spin-wave dispersion relations along the $c$ axis
compared with the experimental results obtained at 2.5 K by Stock {\it et
al.}\cite{Stock} along $(2,0,q)$. The calculated energies of the two $w$ modes
and the two branches of the $c$ mode are shown
by, respectively, the blue and orange lines, and the spectrum includes only
the two times two branches observed in the experiments. The solid lines show
the results obtained when introducing the calculated values of the energy
splittings and the matrix elements in Eqs.\ (\ref{e11}) and (\ref{e13}), whereas
the dashed lines are the results obtained using the linear spin-wave
expressions (\ref{e15}) and (\ref{e16}).}
\label{f4}
\end{figure}

The calculated results for the susceptibility components as functions
of temperature are compared with experiments in Fig.\ \ref{f3}. The
dashed line in this figure shows the MF susceptibility for the same
model given by (\ref{e19}), but with the intra-triangle interaction
$J_1^{}$ included as a mean-field contribution. This simple MF model
leads to a N\'eel temperature which is 70.1 K, whereas the present
cluster-MF model predicts $T_N^{}=36.7$ K in reasonable agreement
with the experimental value $T_N^{}=27$ K. The calculated Curie
temperature is $\theta=-190$ K, and the one derived from the
experimental paramagnetic susceptibility in Fig.\ \ref{f3} is
$\theta=-188$ K (using the data within the whole interval between
$T_N^{}$ and 300 K), hence the experimental frustration factor,
$f=-\theta/T_N^{}$,\cite{Ramirez} is about 7 for this system. Because
of the relatively strong intra-triangle interaction $J_1^{}$, the
system is frustrated, which invalidates the simple MF approximation.
Within the cluster-MF approximation, the spin triangles are treated
as correlated units, implying that the main source for frustration,
$J_1^{}$, is accounted for in an exact way. The results above and the
comparison in Fig.\ \ref{f3} show that the MF method is substantially
improved, when choosing the basis to be the spin triangles instead of
the individual spins.

\begin{figure*}[t]
\includegraphics[width=0.38\linewidth]{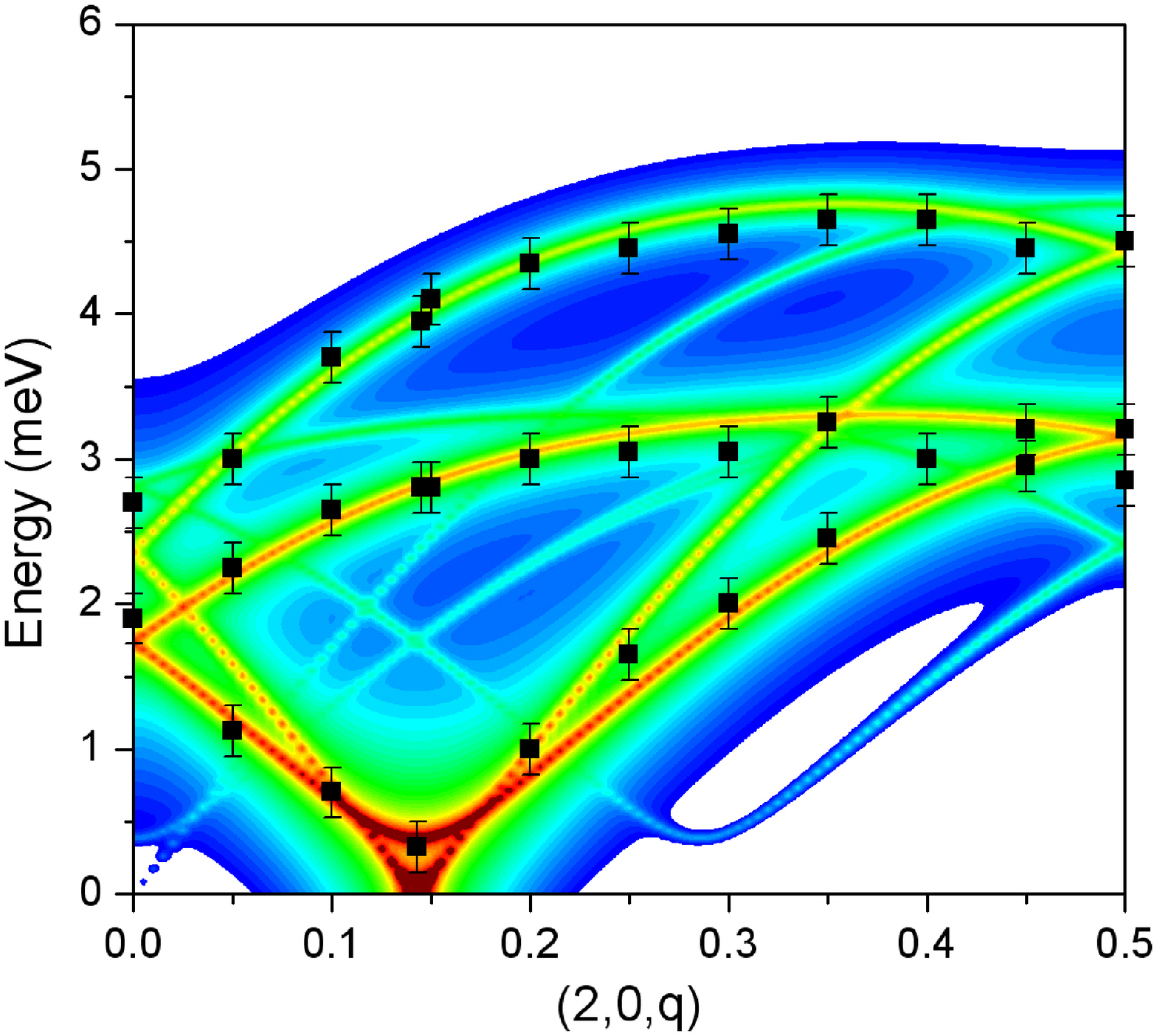}
\includegraphics[width=0.38\linewidth]{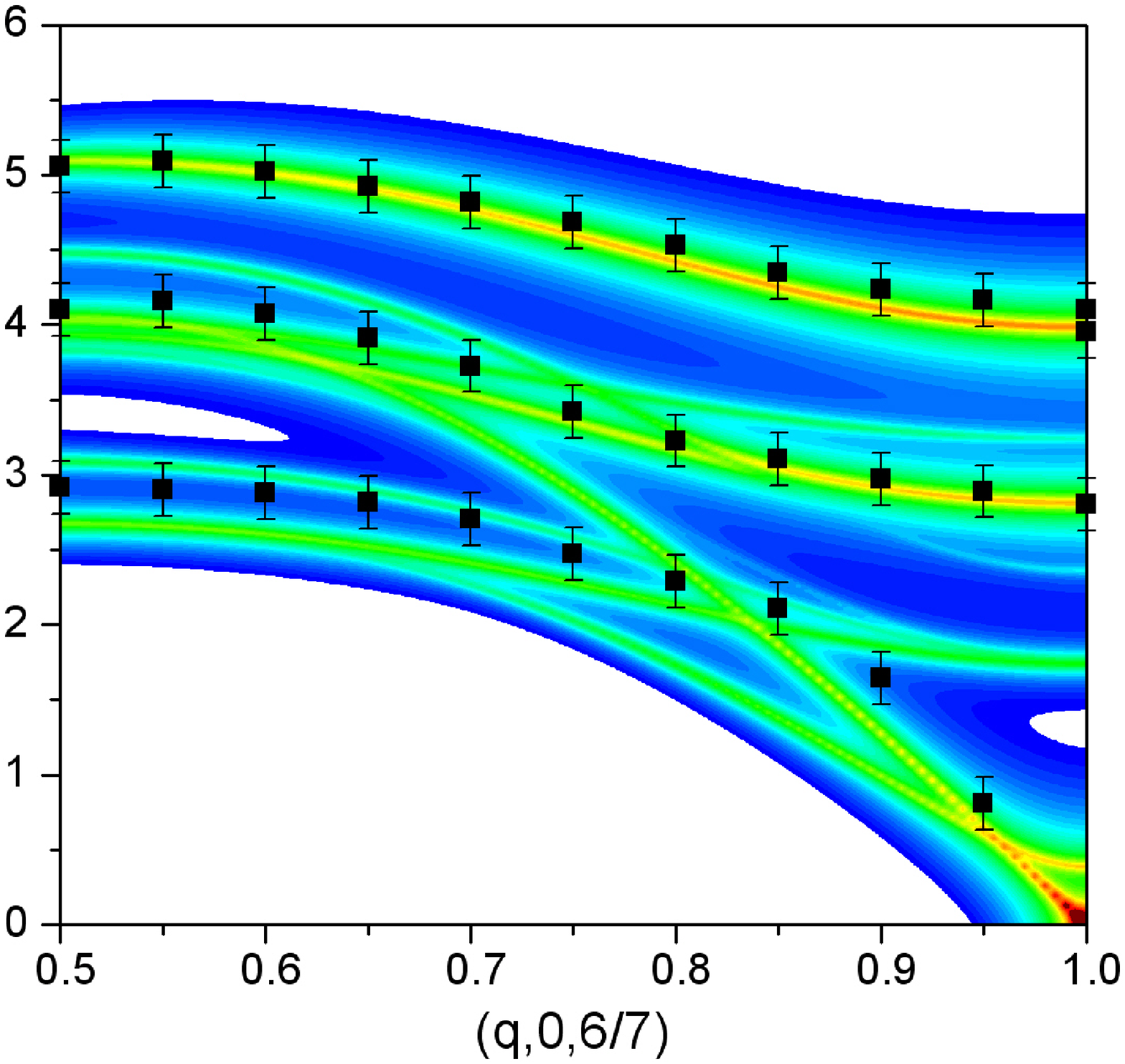}
\includegraphics[width=0.22\linewidth]{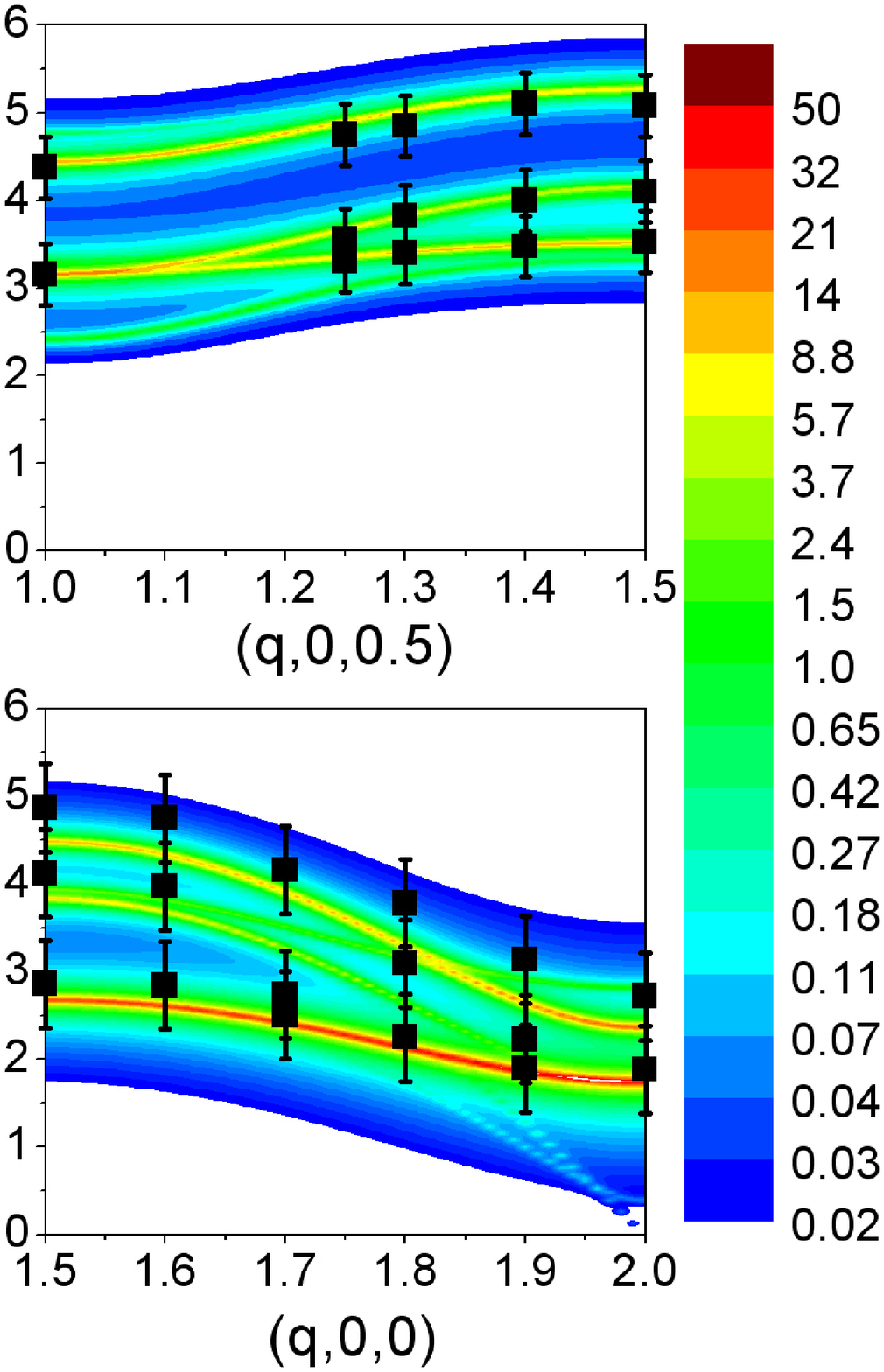}
\caption{(Color online) Logarithmic contour plots of the calculated
spin-wave scattering intensities along different scattering vectors
in reciprocal lattice units. The black squares indicate the
experimental energies determined at $T=2.5$ K by Stock {\it et
al.}\cite{Stock} The logarithmic scale shown to the right is common
for all three cases shown here and also applies to Fig.\ \ref{f7} and
to the left figure in Fig.\ \ref{f8}. The use of the logarithmic
scale may be slightly misleading in the sense that the low intensity
branches (light green/light blue), which are easily identified in
these plots, are probably not visible under realistic circumstances.}
\label{f5}
\end{figure*}

\begin{figure}[t]
\includegraphics[width=0.49\linewidth]{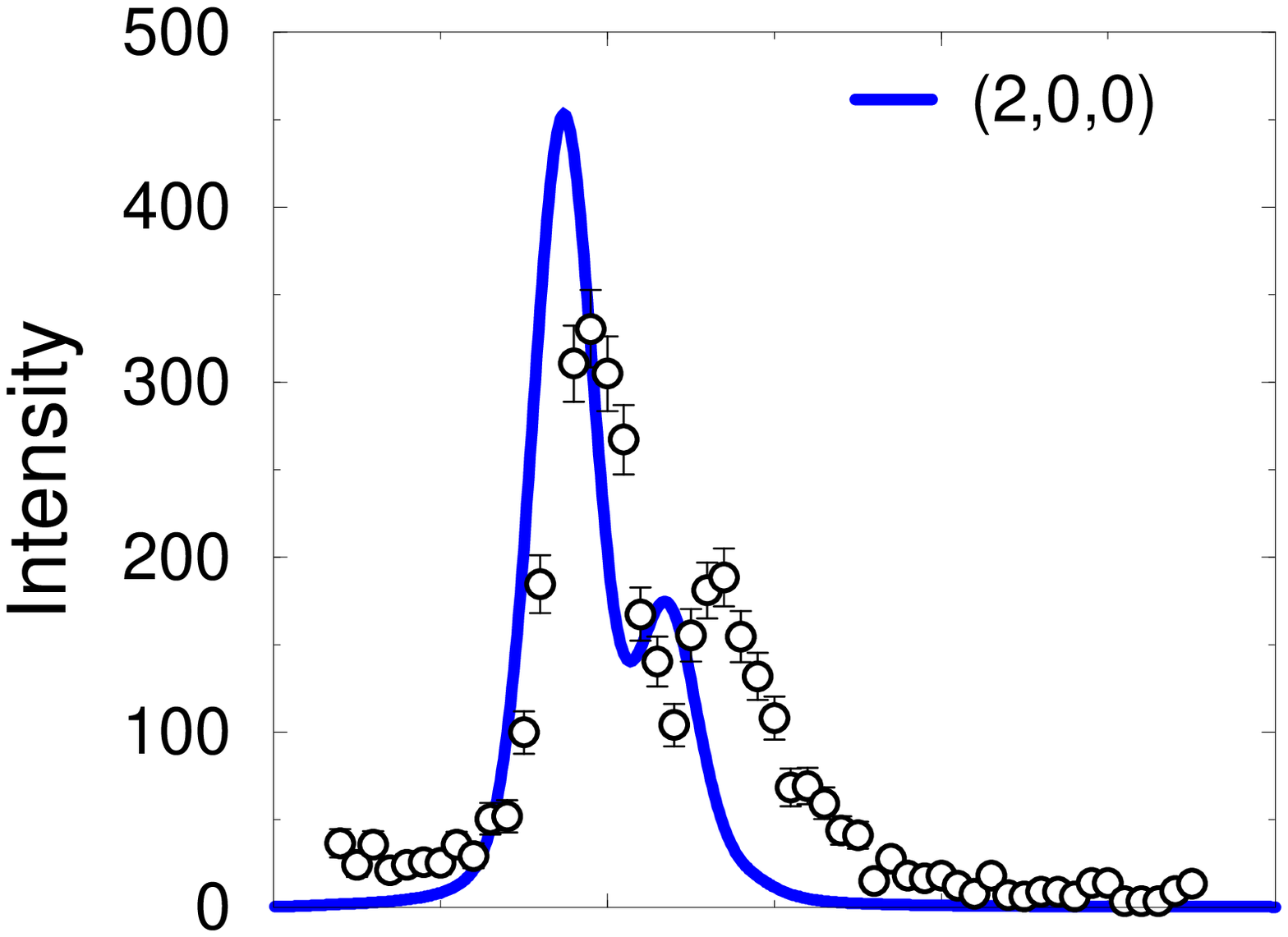}
\includegraphics[width=0.49\linewidth]{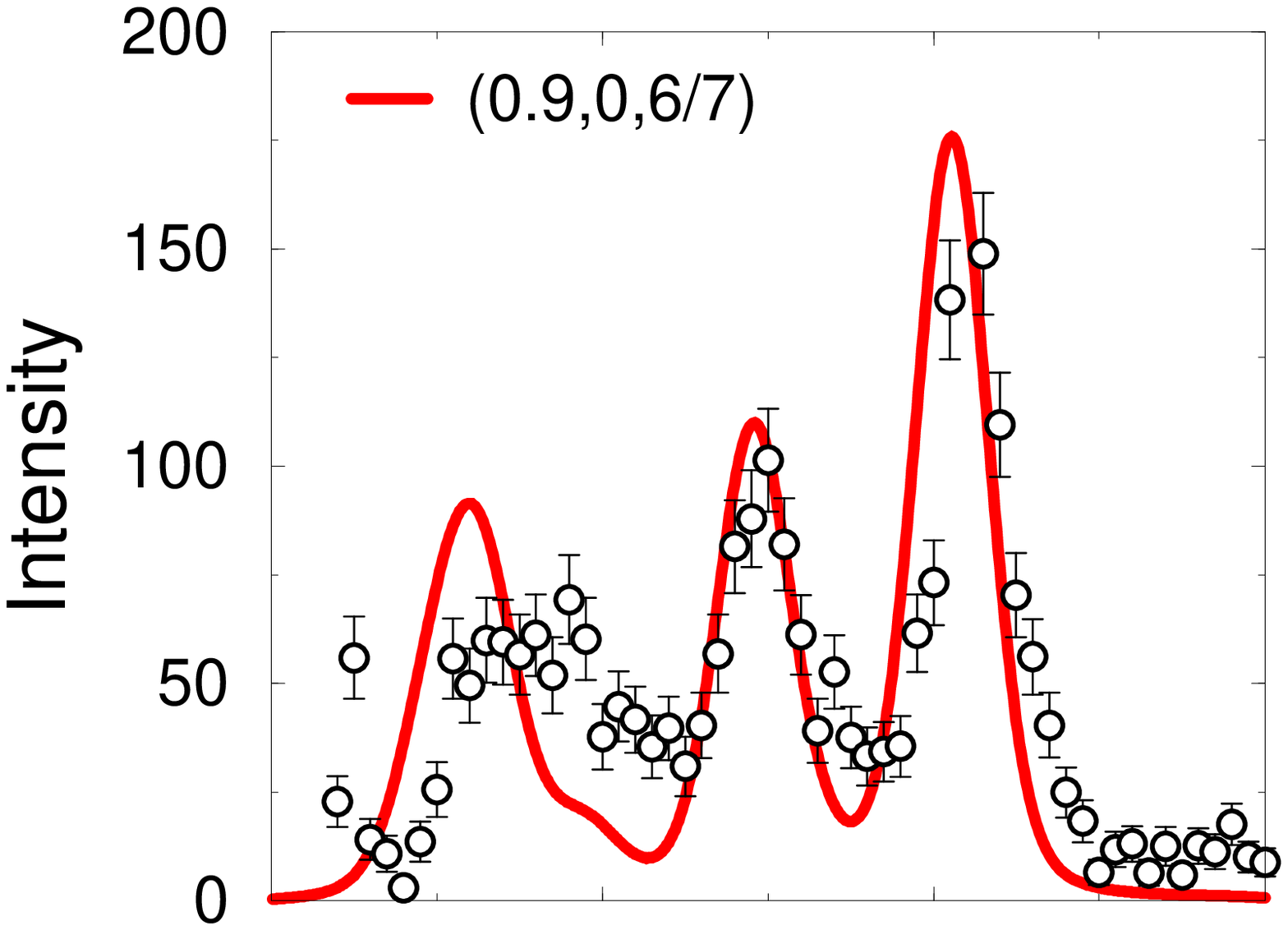}
\includegraphics[width=0.49\linewidth]{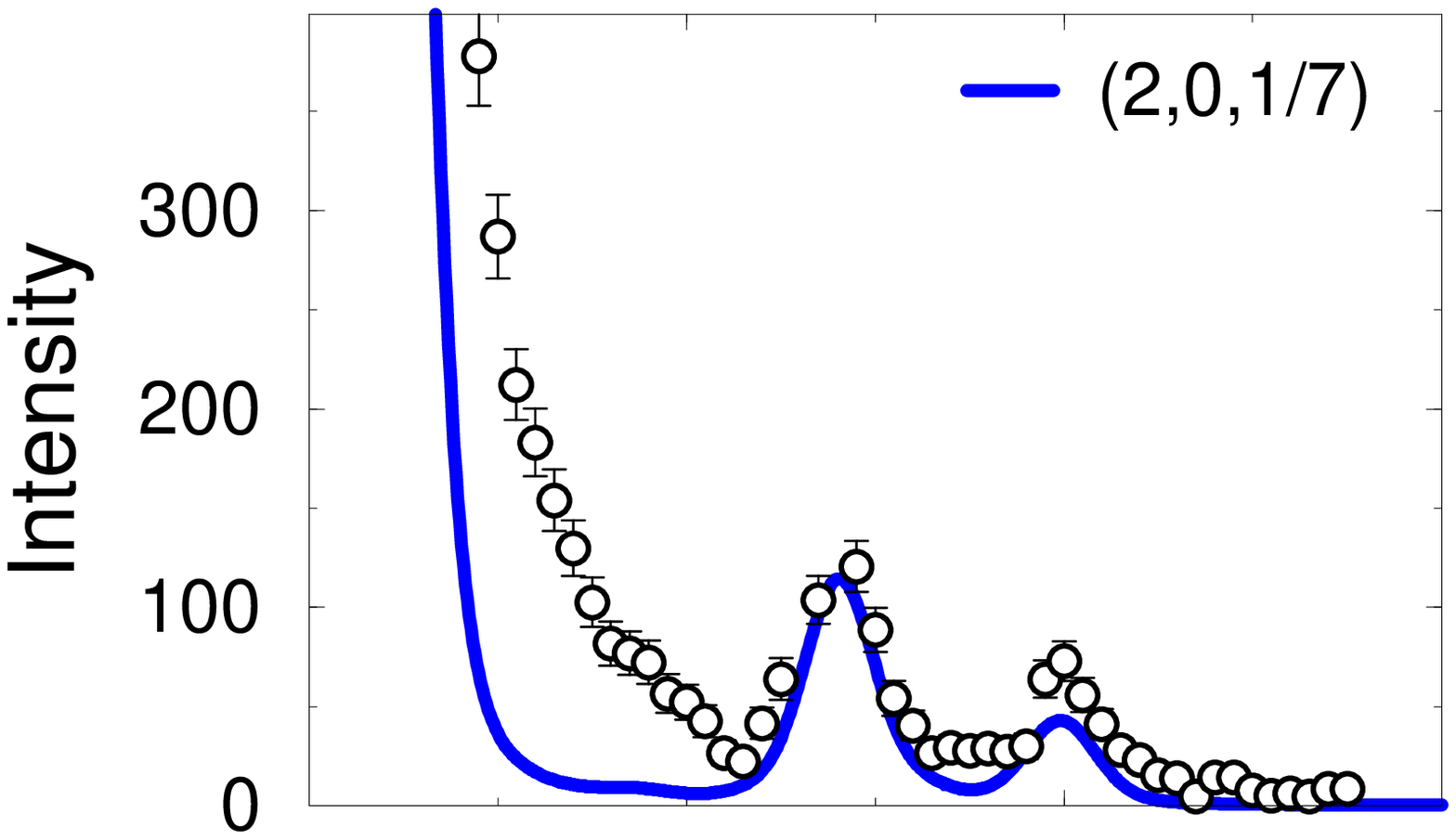}
\includegraphics[width=0.49\linewidth]{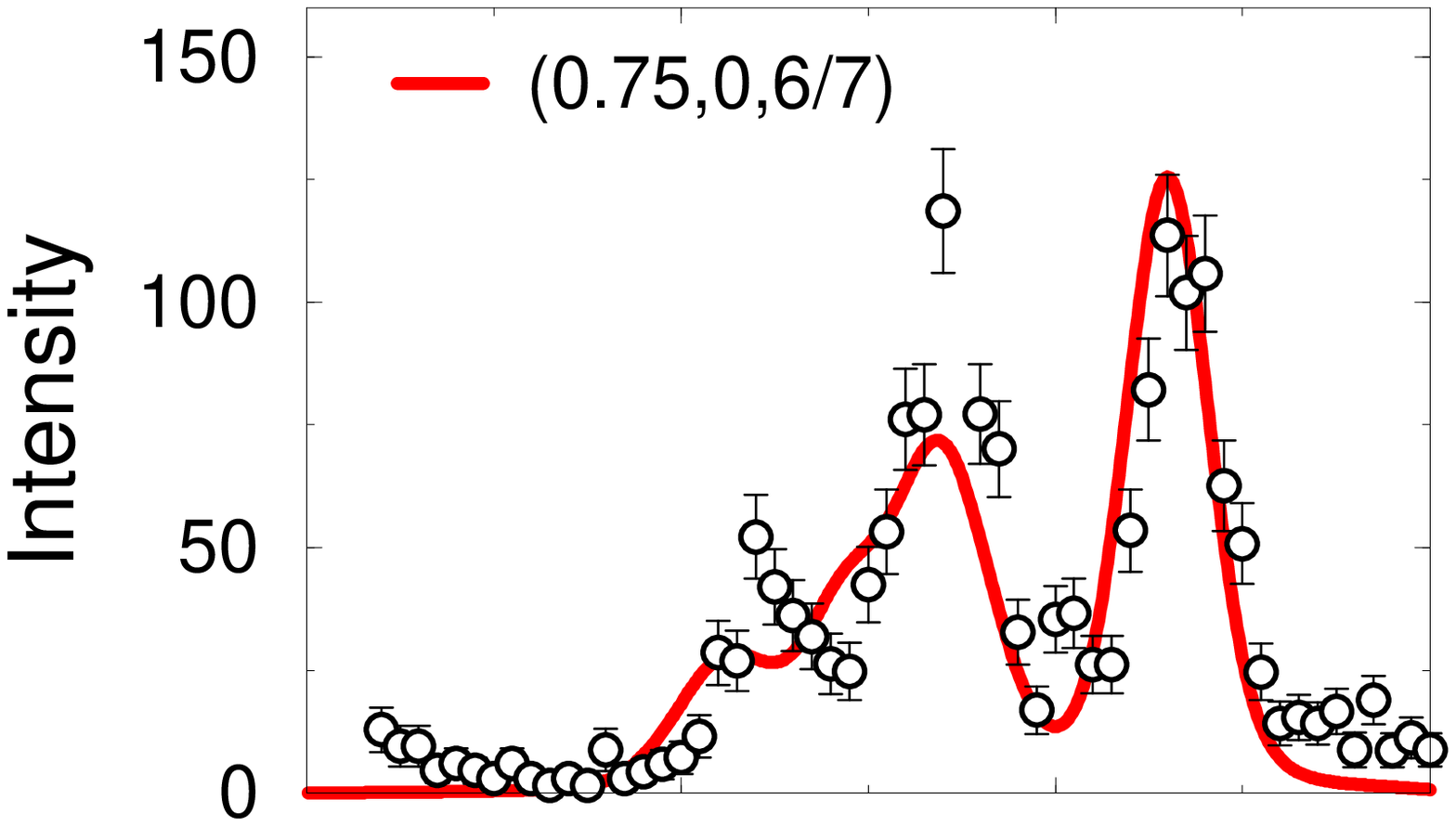}
\includegraphics[width=0.49\linewidth]{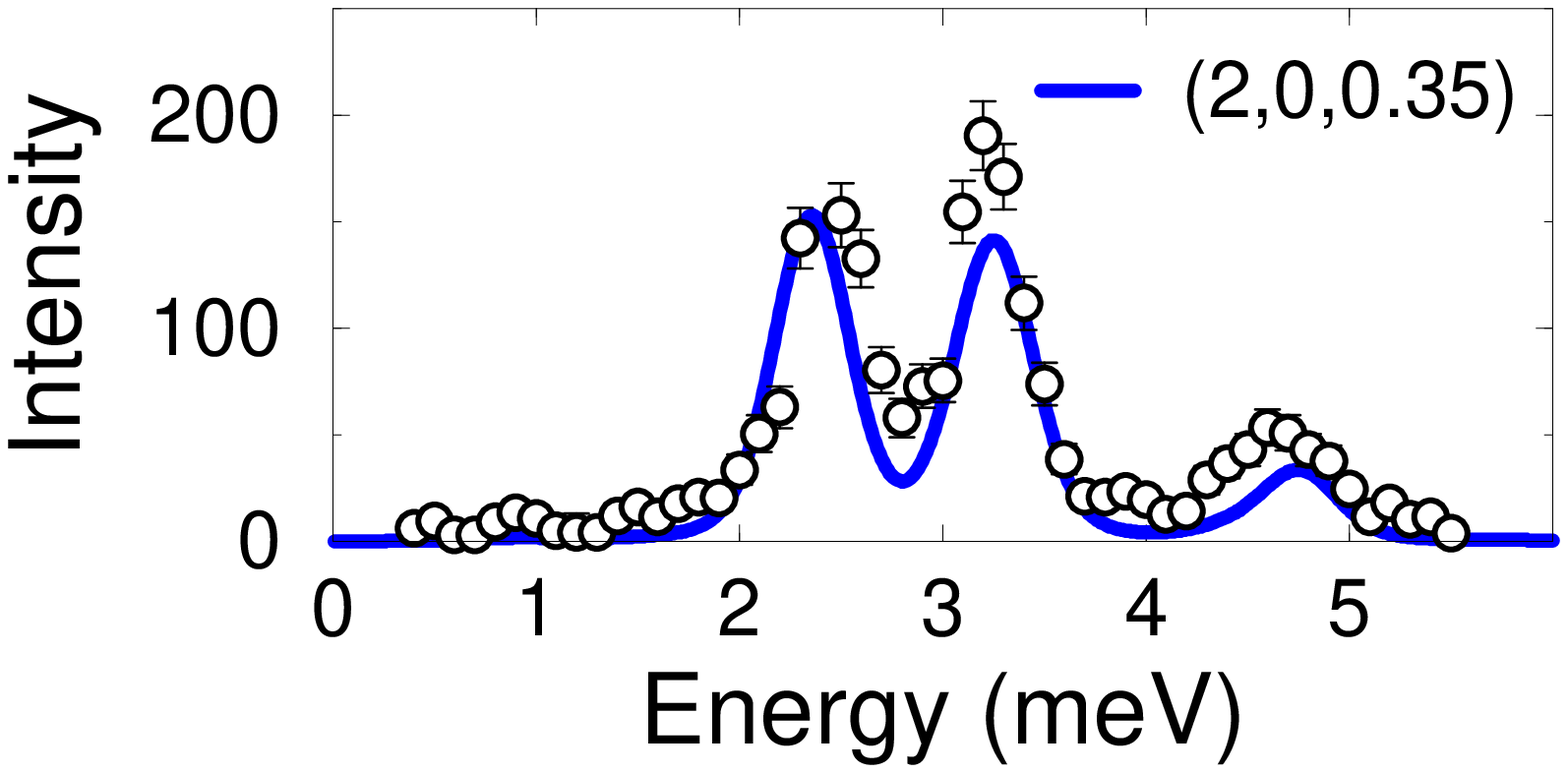}
\includegraphics[width=0.49\linewidth]{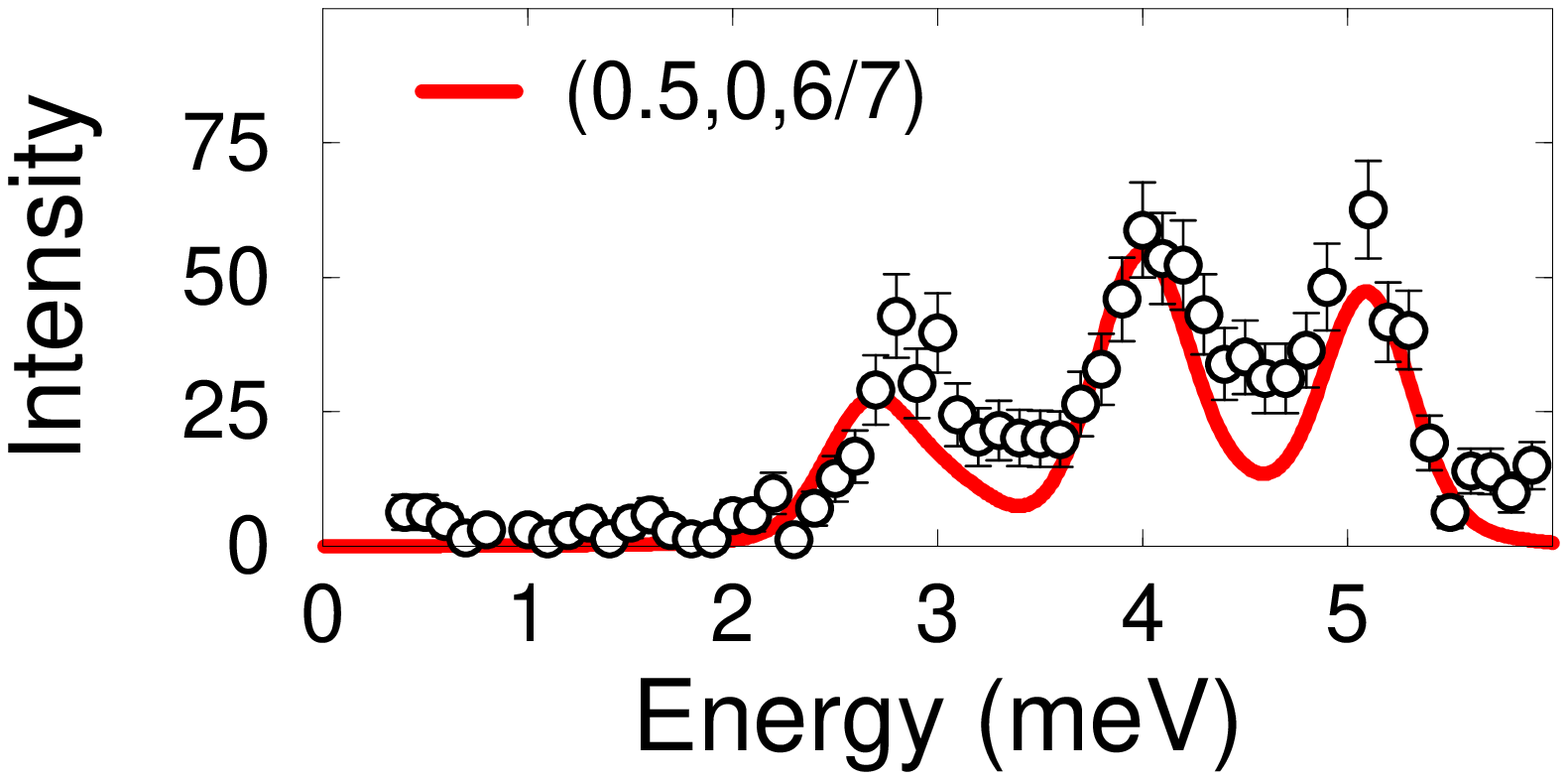}
\caption{(Color online) Detail comparisons between the experimental\cite{Stock}
and calculated cross sections for a selection of the scattering vectors
considered in Fig.\ \ref{f5}.}
\label{f6}
\end{figure}

The parameters given by (\ref{e19}) lead to an exchange interaction
$J(Q)=-0.821$ meV. The self-consistent diagonalization of the
cluster-MF Hamiltonian in (\ref{e1}) with $h=-\langle S_{\xi
z}^{}\rangle J(Q)$ predicts the moment $\langle g\mu_B^{}S_{\xi
z}^{}\rangle$ to be 4.71 $\mu_B^{}/$Fe in the zero temperature limit.
The moment is reduced from its fully polarized value of 5
$\mu_B^{}/$Fe, because $|000\rangle$ is not an eigenstate for the
$J_1^{}$ part of the Hamiltonian (\ref{e1}). Introducing the model
parameters in the expression (\ref{e2}) for the mixing parameter, the
result turns out to be $\lambda=0.57$. Hence, $\lambda$ is not small
compared to 1, and the approximate expressions for the spin wave
energies given by (\ref{e15}) and (\ref{e16}) would not be expected
to apply. Nevertheless, the analytic spin-wave theory leads to useful
results both when applying directly the final results, Eqs.\
(\ref{e15}) and(\ref{e16}), or when using instead the correct cluster
MF-values for the energy differences and matrix elements in the
expressions (\ref{e11}) and (\ref{e13}), as shown by respectively the
dashed and the solid lines in Fig.\ \ref{f4}. The numerical
diagonalization of (1) predicts $\Delta_a^{}=3.486$ meV and
$\Delta_c^{}=4.751$ meV, and the relative matrix elements are found
to be $m_x^c=1.520$, $m_y^c=0.618$, $m_x^a=0.655$, and $m_y^a=1.275$.
The dispersion relations along the $c$ axis shown in Fig.\ \ref{f4}
have been calculated when introducing these values in the spin-wave
expressions (\ref{e11}) and (\ref{e13}). The results are in good
agreement with experiments and with the scattering intensity maxima
obtained numerically. The only difference between the numerical RPA
calculations (see below) and the spin-wave results given by Eqs.\
(\ref{e11}) and (\ref{e13}) and shown by the solid lines in Fig.\
\ref{e4} is that the simplified model neglects the possible
influences of the higher lying MF levels. This minor inaccuracy is
the reason why the calculated $c$-mode shows a small energy gap at
$Q$ and that $E_w^{}(Q)$ is too large in comparison with the
numerical RPA result. The analytical spin-wave theory is valuable not
because it is nearly able to reproduce the numerical results, but
because it allows a precise interpretation of the numerical RPA
calculations. It is possible to extent the analytic theory to cases
where the wave vector also has a component in the $ab$ plane. This is
a more complex situation, because all three levels are being mixed
with each other.\cite{Jano} The most important change is that
$J_2^{}$ in the spin-wave energies are being replaced by
$J_2^{}[\cos(\mathbf{q}\cdot\mathbf{a}) +
\cos(\mathbf{q}\cdot\mathbf{b})
+\cos(\mathbf{q}\cdot\mathbf{a}+\mathbf{q}\cdot\mathbf{b})]/3$. I
have not found it necessary to carry through these more demanding
calculations, and neither have I tried to work out the analytical
results for the spin-wave scattering intensities, because all these
extra complications are handled in a satisfactory and more accurate
way by the numerical RPA calculations.

The numerical RPA results derived for the spin-wave scattering
intensities are presented in Figs.\ \ref{f5} and \ref{f6}. In all the
calculations, the intensity variations due to the magnetic form
factor of the Fe-ions are neglected. The calculations of the
logarithmic intensities are done with a narrow resolution (a
Lorentzian with $\Gamma=0.05$ meV), whereas a Gaussian resolution
width and an intensity scale factor have been used as fitting
parameters in the direct comparisons shown in Fig.\ \ref{f6}. The two
fitting parameters are the same for all the results shown by the blue
lines, and both parameters have been increased by 20\% in the results
shown by the red lines in the right part of Fig.\ \ref{f6}. These
calculated results are compared with the experimental cross sections
obtained by Stock {\it et al.}\cite{Stock}

\begin{figure}[t]
\includegraphics[width=0.95\linewidth]{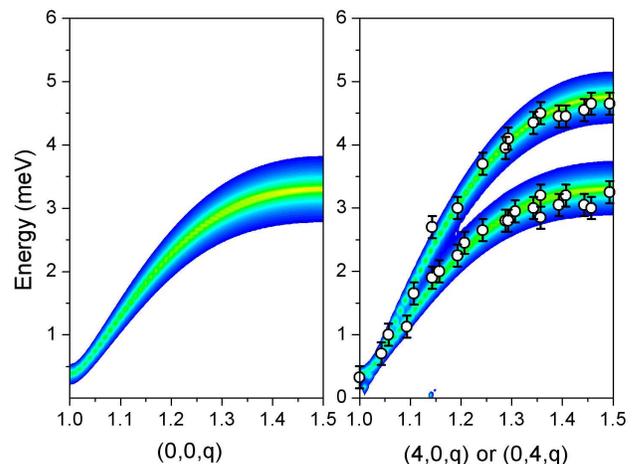}
\caption{(Color online) The logarithmic scattering intensities calculated along $(0,0,q)$
and $(4,0,q)/(0,4,q)$. The experimental results included in the $(4,0,q)/(0,4,q)$
figure are the same as those appearing in Fig.\ \ref{f5} and obtained by
Stock {\it et al.}\cite{Stock}, but the results have been translated by
wave vectors $(0,0,\ell\pm Q)$.}
\label{f7}
\end{figure}

As discussed in the previous section, a scan along $(0,0,q)$ should
show only two nearly degenerate $w$ branches starting out from the
crystallographic Bragg point at $q=1$ (in reciprocal lattice units)
with the energy $E_w^{}(Q)$, and the calculated scattering
intensities obtained for such a scan are shown in Fig.\ \ref{f7}.
Preliminary measurements by Stock {\it et al.}\cite{Stockun} are
consistent with this prediction. The structural parameters determined
by Marty {\it et al.},\cite{Marty} show that the distance $d$ between
the ions in the Fe triangles is very close to be equal to
$(\sqrt{3}/4)\,a$ and thereby that $b^\ast d=\pi$. This means that
the structure factor of the spin triangles is unchanged, if
$4\mathbf{a^\ast}$ or $4\mathbf{b^\ast}$ is added to the scattering
vector. Hence, when neglecting any form factor effects, the only
difference between a $(0,0,q)$ and a $(4,0,q)/(0,4,q)$ scan is that
scattering due to the $c$-axis spin components, which cancels out in
a $(0,0,q)$ scan, contributes at $(4,0,q)/(0,4,q)$. A scan along
$(4,0,q)$ or $(0,4,q)$ is expected to show not only the two $w$-modes
but also the $c$ mode, all starting out from $q=1$ as illustrated by
Fig.\ \ref{f7}. The $w$ modes starting out from the magnetic Bragg
point do (nearly) not appear in this scan, because the total $c$
component of the spin triangles is zero for these modes. The
experimental results shown in the figure are the same ones as
presented in Fig.\ \ref{f5}, but now translated so that they all fall
on the two branches seen in this scan. The energy gap of the $w$
modes, appearing in this plot at $q=1$, has been determined in the
neutron scattering experiments by Stock {\it et al.}\cite{Stock} and
by Loire {\it et al.}\cite{Loire} to lie between 0.35 and 0.4 meV.
This value of the gap is here used for determining the numerical
value of $D_c^{}$. This way of plotting the experimental results also
shows that the energy of the $c$ mode at $(4,0,1+Q)$, the upper peak
in the $(2,0,0)$ scan shown in Fig.\ \ref{f6}, is distinctively
higher than the value suggested by a sinusoidal interpolation of the
other $c$-mode results. The difference might be caused by special
effects related to this particular wave vector $(2,0,0)$, or it may
be an indication of a weak interaction between spins in next-nearest
neighboring layers. The fit to the two dispersion relations may be
improved by including a ferromagnetic interaction
$J_{12}^{}(2c)\simeq-0.01$ meV (between sublattice 1 and 2 at a
distance of $2c$ along the $c$ axis), however, the improvements are
not really significant and this possible modification is abandoned.

The experiments of Stock {\it et al.}\cite{Stock} were all performed
in the $a^\ast c^\ast$ plane, whereas Loire {\it et al.}\cite{Loire}
did choose the $b^\ast c^\ast$ plane as scattering plane, and they
observed a clear asymmetry between the intensities of unpolarized
neutrons scattered at $(0,1,q)$ and $(0,1,-q)$ caused by $J_3^{}$
being different from $J_5^{}$. The branches starting out from zero
energy at $(0,1,\ell-Q)$ were found to be much more intense than the
$(0,1,\ell+Q)$ branches (see also the left figure showing $S(0,1,q)$
in Fig.\ \ref{f8} below). This is in agreement with the theoretical
predictions, if $J_5^{}$ is the dominant interplanar interaction, or,
more accurately formulated, the theory becomes in accord with these
experimental results, if the sign of $R$ in Eq.\ (\ref{e5}) is chosen
to be negative, which choice is already made with the model
parameters introduced by Eq.\ (\ref{e19}). The experiments of Loire
{\it et al.}\cite{Loire} were done on a crystal for which the
structural chirality was determined to be $\epsilon_T^{}=-1$ from the
anomalous part of the x-ray scattering function.\cite{Marty} This
means that the unpolarized neutron experiments of Loire {\it et
al.}\cite{Loire} show that the strongest interplanar interaction
$J_5^{}$ is, as expected,\cite{Stock,Marty} the one determined by the
shortest super-superexchange path. The conclusion that
$\mbox{sign}(R)=\epsilon_T^{}$ and therefore, according to Eq.\
(\ref{en}), $\epsilon_H^{}=\epsilon_T^{}\,\epsilon_\gamma^{}$, or
$\epsilon_\gamma^{}=\epsilon_H^{}\epsilon_T^{}$, is the same one
derived by Marty {\it et al.}\cite{Marty} from their unpolarized
neutron-diffraction experiments. The elastic and the inelastic
unpolarized neutron experiments independently show that the relation
$\epsilon_H^{}=\epsilon_T^{}\,\epsilon_\gamma^{}=-\epsilon_T^{}\,\mbox{sign}(D_c^{})$
applies, but are unable to decide on the helicity of the magnetic
structure $\epsilon_H^{}$ and, thereby, on the sign of $D_c^{}$ or
the orientation of the ordered spin triangles. The unpolarized
neutron scans in the $a^\ast c^\ast$ plane should show the similar
asymmetry except that the high intensity branch is the one starting
out from $(1,0,\ell+Q)$. Preliminary results by Stock {\it et
al.}\cite{Stockun} show a pronounced asymmetry between the scattering
at $(1,0,\ell+Q)$ and $(1,0,\ell-Q)$, but the branch with the largest
intensity is the one emerging from $(1,0,\ell-Q)$, which indicates
that their crystal has $\epsilon_T^{}=+1$, i.e.\ the opposite
structural chirality to that of the crystal investigated by Loire
{\it et al.}\cite{Loire} The coincidence that the distance between
the Fe ions in the triangles is close to $(\sqrt{3}/4)a$ implies that
the asymmetry disappears around Bragg points $(h,k,\ell)$, when for
example $h+k=2$ like in the case considered in Fig.\ \ref{f5}.

\begin{figure*}[t]
\includegraphics[width=0.317\linewidth]{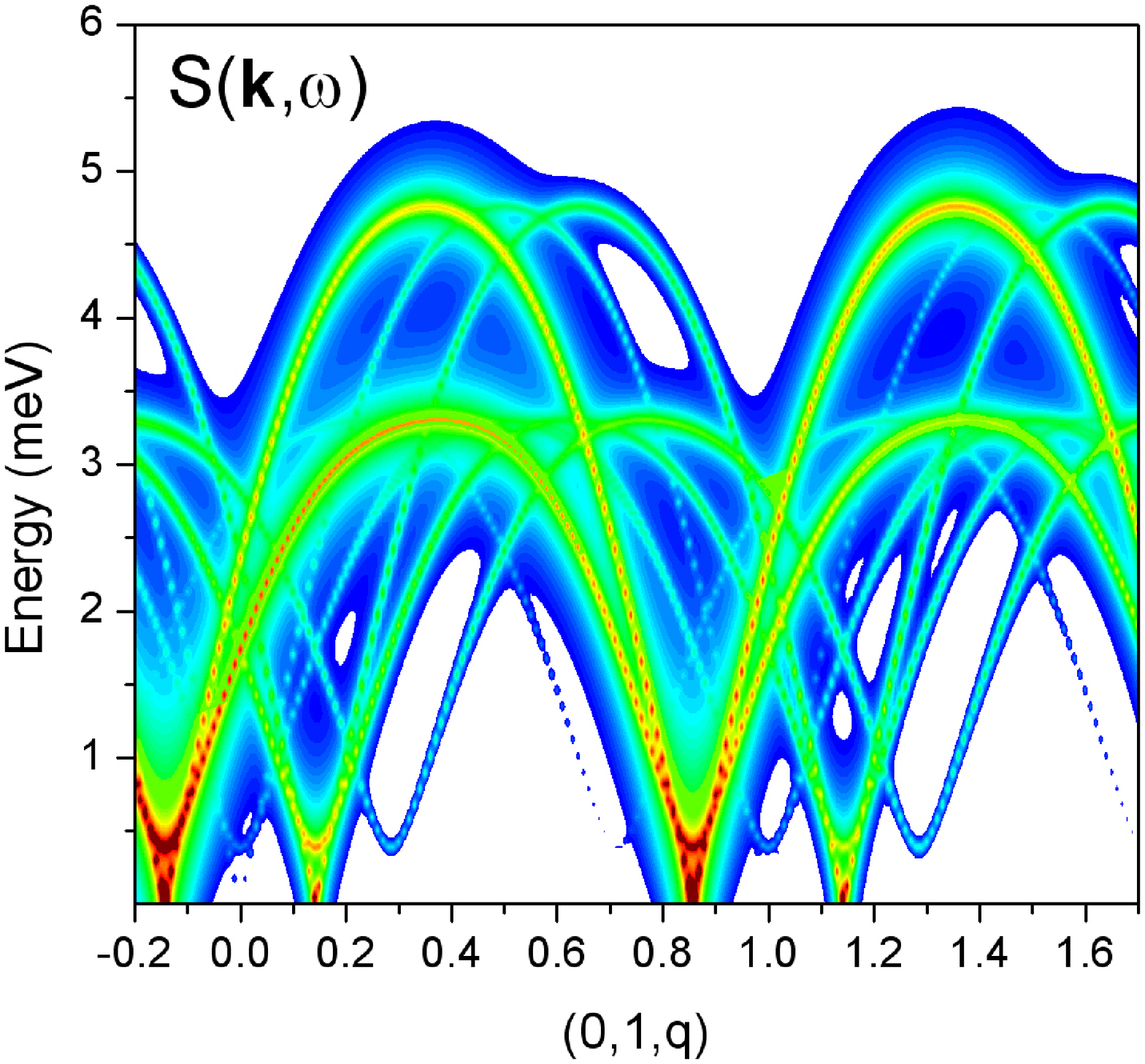}
\includegraphics[width=0.33\linewidth]{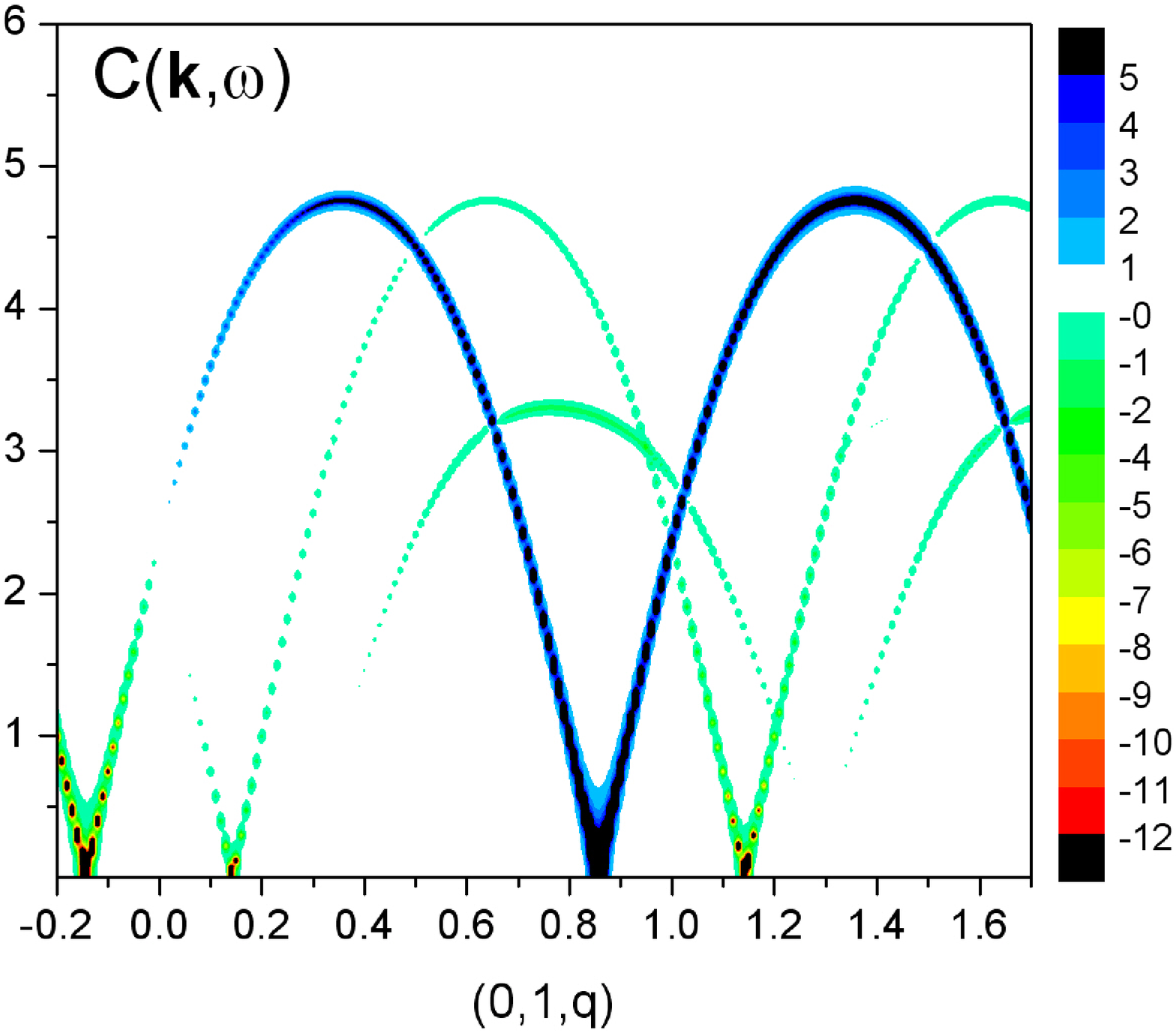}
\includegraphics[width=0.333\linewidth]{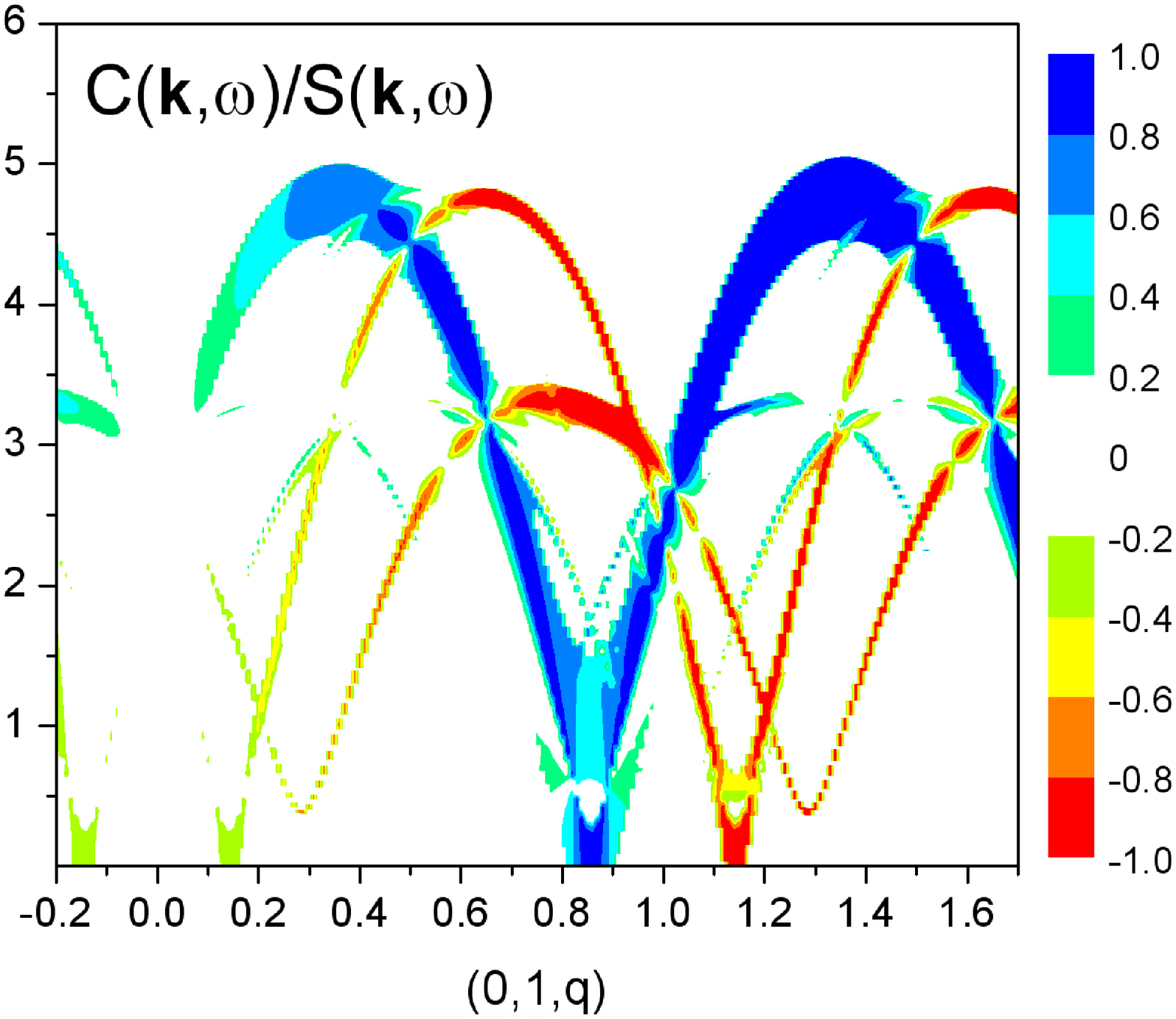}
\caption{(Color online) Contour plots of the calculated spin-wave
scattering intensities along $\mathbf{k}=(0,1,q)$. The left figure
shows the logarithmic intensities $S(\mathbf{k},\omega)$ derived for
unpolarized neutron scattering. Notice that the intensities of the
branches starting out from zero energy at $(0,1,\ell-Q)$ are much
higher than the intensities of the $(0,1,\ell+Q)$ branches. The
figure in the middle shows the calculated polarization function
$C(\mathbf{k},\omega)$ using a linear scale. The figure to the right
shows the corresponding dynamic scattering ratio
$C(\mathbf{k},\omega)/S(\mathbf{k},\omega)$.}\label{f8}
\end{figure*}
\begin{figure}[ht]
\includegraphics[width=0.49\linewidth]{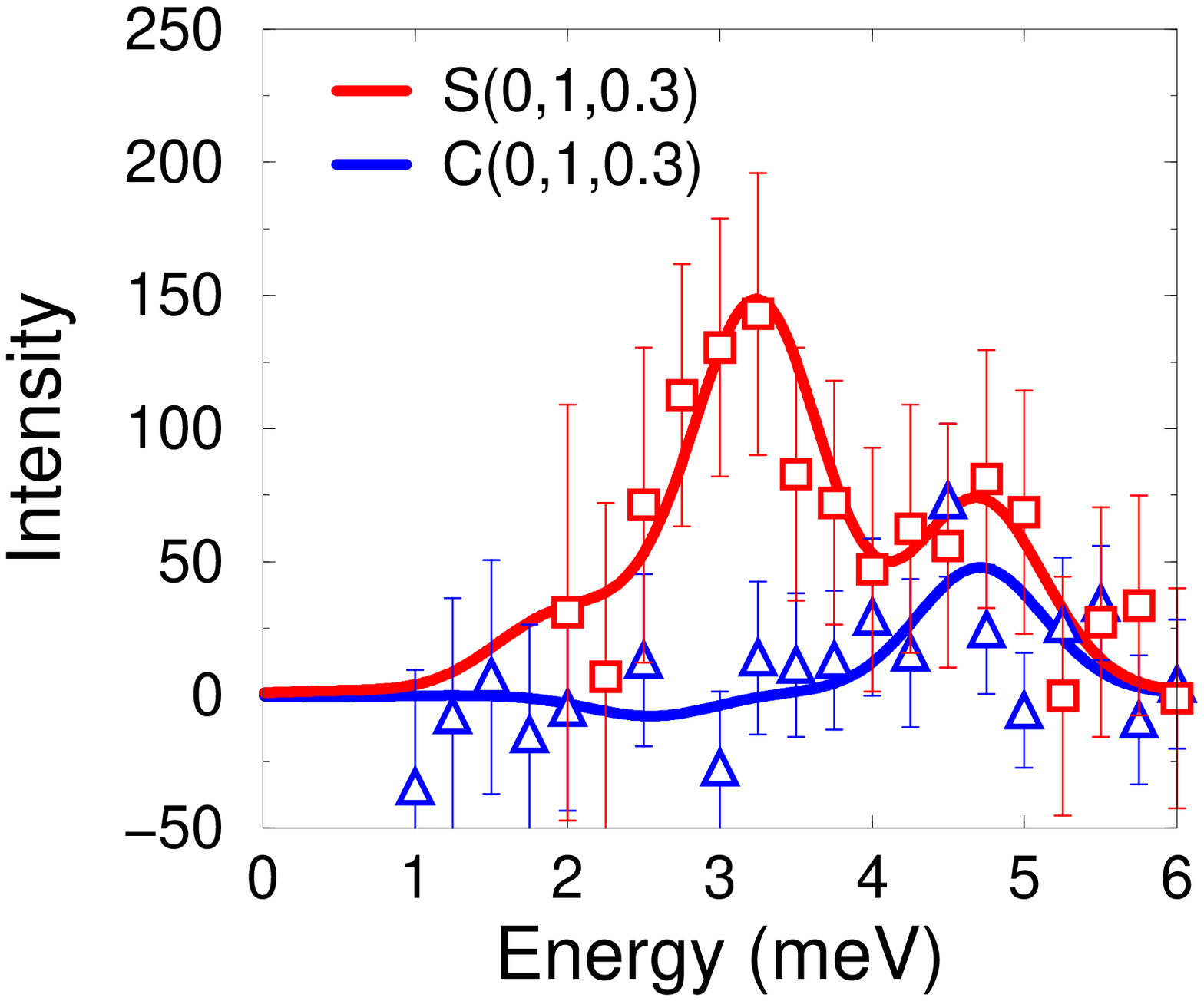}
\includegraphics[width=0.49\linewidth]{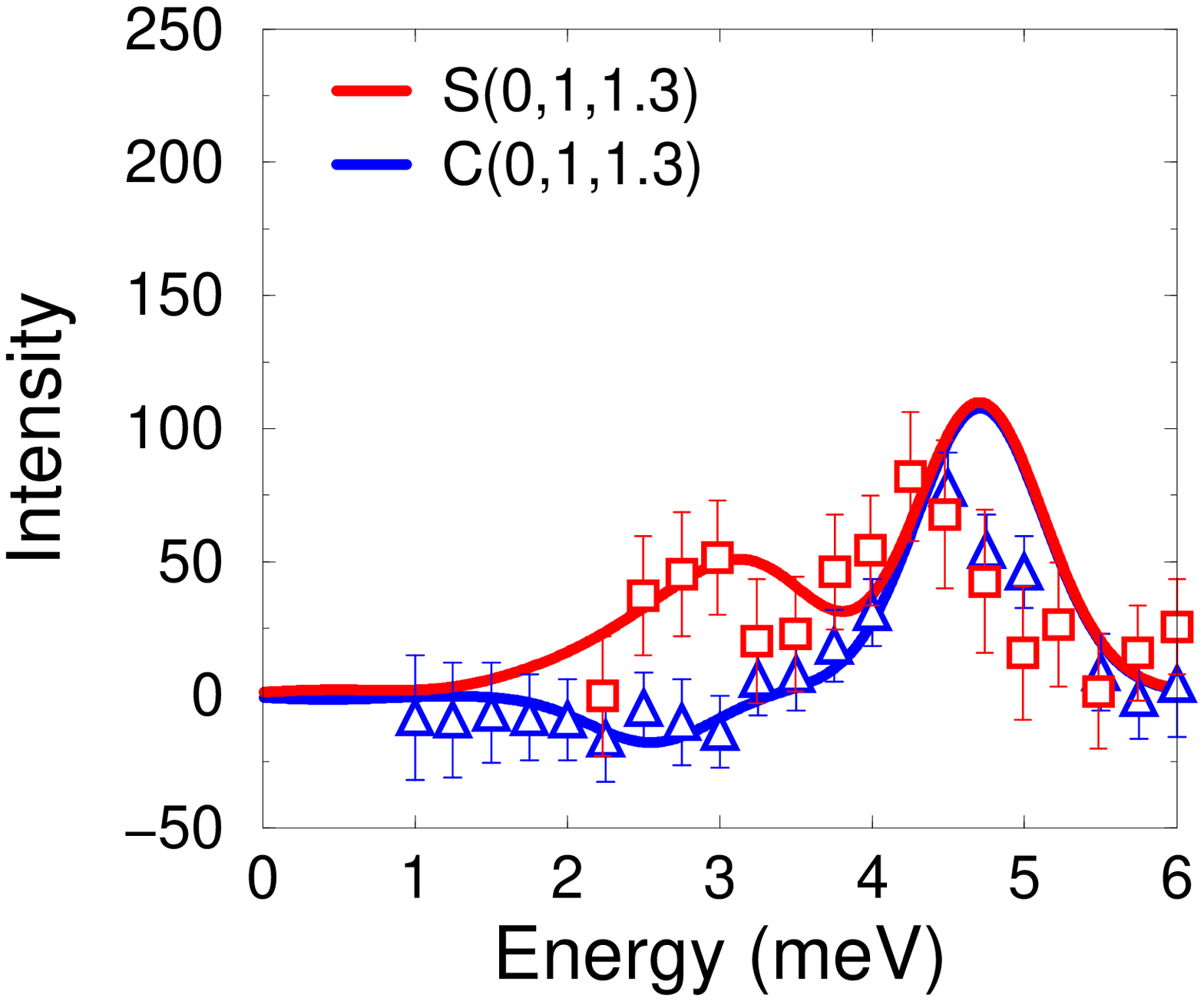}
\caption{(Color online) Comparison between theory and experiments for two of
the scattering vectors considered in Fig.\ \ref{f8}. The experimental results
are obtained by Loire {\it et al.}\cite{Loire} from inelastic scattering of
polarized neutrons. The intensity scaling factor and
the resolution width are kept constant in all the calculated results.}
\label{f9}
\end{figure}

\section{Helically polarized spin waves}
Loire {\it et al.}\cite{Loire} have performed a series of inelastic
scattering experiments in the $b^\ast c^\ast$ plane with polarized
neutrons. In these experiments they determined the spin-flip
scattering intensities $I^{\pm/\mp}_{}$ corresponding to the
differential cross sections (within the terminology of Moon, Riste,
and Koehler\cite{Moon})
\begin{equation}
\frac{d\sigma^{\pm/\mp}_{}}{d\Omega}=\sum_{ij}e^{i\mathbf{k}\cdot\mathbf{r}_{ij}}
\,p_i^{}p_j^\ast\left[\mathbf{S}_{\perp i}^{}\cdot
\mathbf{S}_{\perp j}^{}\mp i \mathbf{\hat{z}}\cdot (\mathbf{S}_{\perp i}^{}
\times\mathbf{S}_{\perp j}^{})\right].\label{e20}
\end{equation}
In their experiments Loire {\it et al.}\ did choose the neutron-spin
polarization vector $\mathbf{\hat{z}}$ to be parallel to the
scattering vector
$\mathbf{k}=\mathbf{k}_{\text{initial}}^{}-\mathbf{k}_{\text{final}}^{}$,
i.e.\ $\mathbf{\hat{z}}=\mathbf{k}/|\mathbf{k}|$, and extracted the
following inelastic scattering functions,\cite{Loire}
\begin{eqnarray}\label{e21}
S(\mathbf{k},\omega)=\frac{I^{\pm}_{}(\mathbf{k},\omega)+
I^{\mp}_{}(\mathbf{k},\omega)}{2}&&\nonumber\\
C(\mathbf{k},\omega)=\frac{I^{\pm}_{}(\mathbf{k},\omega)-
I^{\mp}_{}(\mathbf{k},\omega)}{2}&.&
\end{eqnarray}
In the case where the spins of the incident neutrons are polarized
parallel to the scattering vector $\mathbf{k}$,
$I^{\pm}_{}(\mathbf{k},\omega)$ is the intensity of the scattered
neutrons with the opposite polarization. Since $\mathbf{\hat{z}}$ is
reversed by definition, when $\mathbf{k}$ is replaced by
$-\mathbf{k}$, it is found that
$C(\mathbf{k},\omega)=C(-\mathbf{k},\omega)$ as well as
$S(\mathbf{k},\omega)=S(-\mathbf{k},\omega)$. The expression given by
Eq.\ (16) in Ref.\ \onlinecite{JJ} is straightforwardly generalized
so to include $C(\mathbf{k},\omega)$: the products of spin components
appearing above in Eq.\ (\ref{e20}) are translated into the
corresponding tensor components of the correlation function in Eq.\
(16) of Ref.\ \onlinecite{JJ}, but notice that the sign convention
for the wave vectors applied in this reference is the opposite of
that used above. For a simple righthanded ($\epsilon_H^{}=+1$) helix
with ordering wave vector $Q\mathbf{c^\ast}$ ($0<Q<1/2$), the elastic
response $I^{\mp}_{}$ is non zero, and $I^{\pm}_{}$ is zero, for the
magnetic Bragg peaks at $\mathbf{k}=(0,0,|\ell|+Q)$ and
$(0,0,-|\ell|-Q)$, whereas the opposite is the case for the remaining
Bragg points along $\mathbf{c^\ast}$. The ratio between the static
scattering functions $C(\mathbf{k})$ and $S(\mathbf{k})$ for a simple
helix  with helicity $\epsilon_H^{}$ is in general found to be
\begin{equation}\label{e22}
\frac{C(\mathbf{k})}{S(\mathbf{k})}=\frac{2\cos\theta}{1+\cos^2\theta}
\left[\delta(\mathbf{G}-\mathbf{Q}-\mathbf{k})-
\delta(\mathbf{G}+\mathbf{Q}-\mathbf{k})\right]\,\epsilon_H^{},
 \end{equation}
where $\cos\theta=\mathbf{k}\cdot\mathbf{Q}/|\mathbf{k}||\mathbf{Q}|$
and $\mathbf{G}$ is a reciprocal lattice vector.

The polarized inelastic neutron-scattering experiments performed by
Loire {\it et al.}\cite{Loire} on a crystal with $\epsilon_T^{}=-1$
showed that $C(\mathbf{k},\omega)$ is positive and very nearly equal
to $S(\mathbf{k},\omega)$ at the lowest energies observed (1 meV)
close to the Bragg point at $(0,1,6/7)$, and that this is also the
case for the $c$ mode at $(0,1,1.3)$ as shown in Fig.\ \ref{f9}. The
behavior of $C(\mathbf{k},\omega)$ was measured along $(0,1,q)$, and
the results shown by Loire {\it et al.}\ in their Fig.\ 3(e) are very
similar to the calculated results shown in the middle figure in Fig.\
\ref{f8} (the weak branches included here were not detectable). The
sign of $C(\mathbf{k},\omega)$ determines the sign of the helicity of
the helically ordered moments, and, as concluded by Loire {\it et
al.},\cite{Loire} their results show that the helicity
$\epsilon_H^{}=+1$. This result, in combination with the unpolarized
result $\epsilon_T^{}=\mbox{sign}(R)=-1$ and the maximization of
${\cal J}(Q)$, Eq.\ (\ref{en}), implies that $\epsilon_\gamma^{}=-1$
corresponding to a positive value of $D_c^{}$.

For a simple helix, the spin-wave theory\cite{Maleev,Lovesey}
predicts the dynamic ratio between the scattering functions
$C(\mathbf{k},\omega)/S(\mathbf{k},\omega)$ to be a purely geometric
quantity, like the static one in Eq.\ (\ref{e22}),
\begin{equation}\label{e23}
\frac{C(\mathbf{k},\omega)}{S(\mathbf{k},\omega)}=
\pm\frac{2\cos\theta}{1+\cos^2\theta}
\,\epsilon_H^{}\,.
\end{equation}
This expression applies to the modes emerging from the magnetic Bragg
points, which scatter the neutrons via the planar spin components.
The sign to be chosen for the $\pm$ sign in front is the same as the
sign in front of the corresponding delta function in the magnetic
Bragg scattering ratio in Eq.\ (\ref{e22}), i.e.\ the + sign applies
to the modes starting out from the Bragg points at
$\mathbf{G}-\mathbf{Q}$. In contrast to this, the cross section
deriving from the $c$ component of the spins are not affected by the
helicity, and $C(\mathbf{k},\omega)=0$ for the branches emerging from
the nuclear Bragg peaks. With the restriction that we shall only
consider spin waves propagating along the $c$ axis, this simple
description applies almost unchanged to the present ordered system
consisting of three sublattices of coupled helices.
$C(\mathbf{k},\omega)=0$ for the $c$ modes emerging from the nuclear
Bragg peaks at $\mathbf{G}$ and for the $w$ modes starting out from
the magnetic Bragg peaks at $\mathbf{G}\pm\mathbf{Q}$. The $c$ modes
detected via the spin components in the $ab$ plane, are the branches
emerging from the magnetic Bragg points $\mathbf{G}\pm\mathbf{Q}$,
and the chiral scattering ratio for these modes is determined by Eq.\
(\ref{e23}). In principle, the $w$ modes starting out from a nuclear
Bragg point or from $\mathbf{G}\pm2\mathbf{Q}$ should show a dynamic
scattering ratio which is also determined by Eq.\ (\ref{e23}).
However, the $\pm2\mathbf{Q}$ branches are weak, and the two $w$
modes emerging from the same Bragg point are not easy to separate,
and since the chiral scattering ratios have opposite sign for the two
branches, they are going to appear like a single mode with a chiral
polarization factor close to zero. In the scan along $(0,0,q)$ shown
in Fig.\ \ref{f7}, the upper one of the two nearly degenerate $w$
modes, the $-Q$ branch, has
$C(\mathbf{k},\omega)/S(\mathbf{k},\omega)=+1$ whereas
$C(\mathbf{k},\omega)/S(\mathbf{k},\omega)=-1$ for the other $+Q$
mode. These results are valid in the case where the spin waves are
propagating along the $c$ axis, and they agree in most details with
the dynamic scattering ratios calculated numerically using the RPA
model. One example is shown in the right of Fig.\ \ref{f8}. In
principle, the colors in this figure should become more and more blue
[$C(\mathbf{k},\omega)/S(\mathbf{k},\omega)=+1$] or more and more red
[$C(\mathbf{k},\omega)/S(\mathbf{k},\omega)=-1$] for increasing
values of $q$, however, this systematic behavior is going to be
disturbed whenever intensities from modes with different helicity
factors overlap each other (notice that
$|C(\mathbf{k},\omega)/S(\mathbf{k},\omega)|$ is predicted to be
0.967, nearly 1, already at $q=6/7$). The inelastic cross section is
readily calculated numerically in the general case, but the result
becomes less transparent due to the complication that the $c$ and $w$
modes are coupled whenever the propagation vector has a non-zero
component in the $ab$ plane.

\section{Conclusion}
The intra-triangular interaction $J_1^{}>0$ is the dominant cause for
the relatively strong frustration shown by the present spin system,
and it is important to account for this interaction primary to the
interactions between the Fe triangles. The MF approximation, when
applied to the trimerized clusters, leads to a much improved
description of the system in comparison with the single-spin
approximation. The susceptibility just above $T_N^{}$ is being
reduced by 15\% and the transition temperature by 50\%. The more
precise description of the ground state is essential for the
determination of the exchange constants. The analysis of the spin
waves in terms of the five exchange constants, plus the DM
anisotropy, leaves one exchange constant as a nearly free variable.
In the present analysis this degree of freedom has been fixed by a
fitting to the susceptibility. The present model accounts very well
for the susceptibility, whereas the use of, for instance, the
exchange parameters proposed by Loire {\it et al.}\cite{Loire} leads
to values for the susceptibility, which are about 25\% larger than
the experimental ones (40\% in the single-spin MF approximation) at
temperatures close to $T_N^{}$. The MF-cluster calculation also
predicts a quantum reduction of the ordered moment, from 5 to 4.7
$\mu_B^{}$. This reduction is less pronounced than the one derived
from experiments, as the value of the ordered moment determined from
neutron-diffraction experiments at 2 K by Marty {\it et
al.}\cite{Marty} was found to be as small as about 4 $\mu_B^{}$.

Although the trimerized states of the Fe spin triangles differ greatly
from that determined by fully polarized spins, the linear spin-wave theory
works surprisingly well demonstrating the effectiveness of the boson
representation obtained by, for instance, the Holstein--Primakoff
transformation. This advantageous property is of similar significance for
the analysis of an anisotropic ferromagnet like terbium metal, which is
discussed at length in Refs.\ \onlinecite{REM,JJ75}. The particular model
derived by Loire {\it et al.}\cite{Loire} on the basis of linear spin-wave
theory is questioned, but their theoretical results are found to agree
closely with the predictions of the present theory. Their characterization
of the different spin-wave modes may be imprecise, but their main
conclusions are the same as derived here. In the crystal they investigated
the crystallographic chirality is $\epsilon_T^{}=-1$, the ordered moments
in the triangles have the orientation specified by
$\epsilon_\gamma^{}=-1$, and the spiraling ordered moments along a line
parallel to the $c$ axis are making a righthanded helix corresponding to
$\epsilon_H^{}=+1$.

The mirroring of the present system with respect to, for instance,
the $ac$ plane through $\mathbf{S}_1^{}$ would change the structural
chirality $\epsilon_T^{}$ from $-1$ to $+1$, and the helicity
$\epsilon_H^{}$ from $+1$ to $-1$, whereas the sign of $D_c^{}(>0)$
and $\epsilon_\gamma^{}=-1$ would remain unchanged. The choice
between the two possible orientations of the ordered spin triangles
is determined by the local surroundings not by the structural
chirality. By definition chirality should be invariant with respect
to time reversal but have odd parity with respect to
inversion.\cite{Flack} Based on this definition, the orientation of
the spin triangles denoted by $\epsilon_\gamma^{}$ cannot be
characterized as a chiral property.

To summarize: The asymmetry between the inelastic intensities of
magnetic scattered unpolarized neutrons at $(1,0,\ell-q)$ and
$(0,1,\ell+q)$ or at $(1,0,\ell-q)$ and $(1,0,\ell+q)$, as observed,
respectively, by Loire {\it et al.}\cite{Loire} and by Stock {\it et
al.}\cite{Stockun}, is alone a consequence of the structural
chirality. $\epsilon_T$ is equal to $-1$ in the crystal studied by
Loire {\it et al.}, whereas $\epsilon_T^{}=+1$ is the preliminary
result for the crystal investigated by Stock {\it et al.} The DM
anisotropy term $D_c^{}$ in Eq.\ (\ref{e01}), due to the spin-orbit
interaction, couples chirality in spin space to the crystallographic
chirality, $\epsilon_H^{}=\epsilon_\gamma^{}\,\epsilon_T^{}$. The
sign of the DM anisotropy determines the choice between the two
possible orientations of the ordered moments in the Fe triangles, and
it is found that $\epsilon_\gamma^{}=-\mbox{sign}(D_c^{})=-1$. The
combination of an enantiopure crystal and a non-zero DM anisotropy
implies that the system only contains ordered moments with a single
sense of helicity, that $\epsilon_H^{}=-\epsilon_T^{}$ in this system
where $D_c^{}$ is positive. The dynamics of the present system is
unique, not because the spin waves have chiral properties but because
the presence of only one domain of helicity has made it possible to
observed this intrinsic dynamic property. It may be concluded that
the three different modes of spin waves propagating along the $c$
axis in the Ba$_3$NbFe$_3$Si$_2$O$_{14}$ crystal investigated by
Loire {\it et al.}\cite{Loire} should all posses, depending on their
effective propagation vector $\mathbf{q}=\mathbf{k}-\mathbf{G}$, the
same or the opposite sense of helicity as the ordered structure. The
observed behavior of the $c$ mode is in agreement with this
conclusion, whereas the near degeneracy of the $w$ modes starting out
from the nuclear Bragg points prevents an experimental determination
of the chiral properties of these modes.

\begin{acknowledgments}
Des McMorrow is gratefully acknowledged for stimulating discussions
and for providing me with the neutron scattering results obtained by
Stock {\it et al.}\cite{Stock,Stockun} prior to publication. I also
want to thank Stephen Lovesey for making me aware of the precise
definition of ``chirality''.
\end{acknowledgments}

\end{document}